\documentclass{article}
\usepackage[nonatbib, final]{neurips_data_2021}
\usepackage[numbers]{natbib}
\usepackage[utf8]{inputenc} % allow utf-8 input
\usepackage[T1]{fontenc}    % use 8-bit T1 fonts
\usepackage{hyperref}       % hyperlinks
\usepackage{url}            % simple URL typesetting
\usepackage{booktabs}       % professional-quality tables
\usepackage{amsmath}
\usepackage{amsfonts}       % blackboard math symbols
\usepackage{nicefrac}       % compact symbols for 1/2, etc.
\usepackage{microtype}      % microtypography
\usepackage{xcolor}         % colors
\usepackage{soul}
\usepackage{bm}
\usepackage{cleveref}
\usepackage{upgreek}
\usepackage{mathrsfs}  
\usepackage{url}
\usepackage{subfig}
\usepackage{amsthm}
\usepackage{caption}
\newtheorem{definition}{Definition}
\usepackage[inline]{enumitem}
\usepackage{makecell}
\usepackage{boldline}
\setcellgapes{3pt}

\usepackage{todonotes}
\newcommand{\Note}[2]{}
\renewcommand{\Note}[2]{\todo[color=#1,size=\small]{#2}}
\title{CSFCube -- A Test Collection of Computer Science Research Articles for Faceted Query by Example}

\author{%
  Sheshera Mysore\textsuperscript{1}
  \quad
  Tim O'Gorman\textsuperscript{2}\thanks{Work done while at the University of Massachusetts, Amherst.}
  \quad
  Andrew McCallum\textsuperscript{1}
  \quad
  Hamed Zamani\textsuperscript{1}\\[3pt]
  \texttt{\{smysore, mccallum, zamani\}@cs.umass.edu}\\[3pt]
  \textsuperscript{1}University of Massachusetts, Amherst, \textsc{ma, usa}\\
  \textsuperscript{2}Thorn, \textsc{ca, usa}
}

\begin{document}

\maketitle

\begin{abstract}
Query by Example is a well-known information retrieval task in which a document
is chosen by the user as the search query and the goal is to retrieve relevant
documents from a large collection. However, a document often covers multiple
aspects of a topic. To address this scenario we introduce the task of \emph{faceted Query
by Example} in which users can also specify a finer grained aspect in 
addition to the input query
document. We focus on the application of this task in scientific literature
search. We envision models which are able to retrieve scientific papers
analogous to a query scientific paper along specifically chosen rhetorical
structure elements as one solution to this problem. In this work, the
rhetorical structure elements, which we refer to as \emph{facets},  indicate
objectives, methods, or results of a scientific paper. We 
introduce and describe an expert annotated test collection to
evaluate models trained to perform this task. Our test collection consists of a
diverse set of 50 query documents in English, drawn from computational linguistics and
machine learning venues. We carefully follow the annotation guideline used by
TREC for depth-k pooling (k = 100 or 250) and the resulting data collection
consists of graded relevance scores with high annotation agreement. State of the art
models evaluated on our dataset show a significant gap to be closed in further work.
Our dataset may be accessed here: \url{https://github.com/iesl/CSFCube}
\end{abstract}

\section{Introduction}
\label{sec-intro}   
%\begin{figure}[t]
%     \centering
%     \includegraphics[width=0.5\textwidth]{figures/ex-fqbe.png}
%     \caption{An example query abstract and facet: Here querying with the
%\texttt{background} facet for the paper ``Naturalizing a Programming Language
%via Interactive Learning'' would retrieve ``NLyze: Interactive Programming by
%Natural Language for Spreadsheet Data Analysis and Manipulation'' as one of the
%top retrievals.}
%     \label{fig-fqbe-example}
% \end{figure}

% \input{figures/facetexamples}
The dominant paradigm of information retrieval is to treat queries and
documents as different kinds of objects, e.g., in keyword search. This
paradigm, however, does not lend itself to
exploratory search tasks. On the other hand, paradigms of search such 
Query by Example (QBE) which treat queries and documents as similar kinds of
objects have been considered more suited to exploratory search tasks
\citep{ksikes2014towards, Dimitriadou2014EBE, lissandrini2019exampletutorial}.
QBE has also been used more recently in information extraction for NLP
\citep{Shlain2020spike, Sarwar2020QBEEvent}, and seems poised to leverage recent
advances in representation learning and NLP for search tasks where keyword
search proves insufficient \citep{arguello2021tip, zhu2014multiqbe}.

\begin{figure*}[htp]
     \centering
     \includegraphics[width=\textwidth]{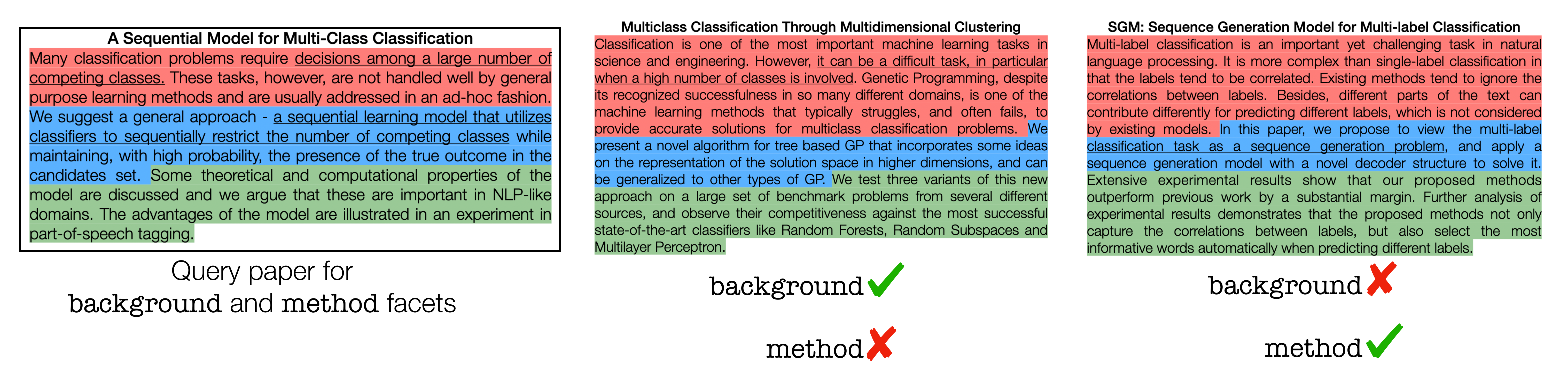}
     \caption{Examples of candidate papers similar along different facets
for the same query paper, underlined text indicates similar aspects: A paper discussing a classification problem with large number of classes is similar along \texttt{background} with a paper discussing the same problem, but with a different paper discussing a sequential model for its \texttt{method} facet.}
     \label{fig:facetexaamples}
 \end{figure*}

In QBE search tasks, each query is a document that often covers
multiple aspects of a topic leading to a document-only query under-specifying how retrievals should me made. Here, we introduce the task of faceted QBE,
where users can specify an information need by providing an input
document and a facet example, with the goal to retrieve documents that are similar to the input document from the perspective of the given facet. This paper focuses on the case of faceted QBE applied to scientific articles.
Figure \ref{fig:facetexaamples} illustrates how multi-aspect similarities may arise in scientific articles, where candidate documents could be similar to the query along the general problem being addressed or the method used in a paper.

The promise of literature search to navigate vast collections of research documents offers exciting prospects and
the potential to accelerate the process of scientific discovery
\citep{GilScienceReview}. This optimism has been expressed in a range of
work in biomedicine \citep{treccovid2021}, materials science
\citep{Himanen2019MSChallenges}, geography \citep{lafia2020designing},
biomimetic design \citep{Kruiper2016biomimetics},
and machine learning \citep{Vanschoren2013OpenML}.
The ever-growing research
literature has also lead to search and recommendation tools 
being important in the workflows of individual researchers across a range of disciplines \citep{landhuis2016infoverload, pain2016keep}. However, few literature search
and recommendation tools address the problem that documents often contain
multiple fine grained aspects one might want to search with \citep{ostendorff2020contextual}. Further, prior work has also
demonstrated that researchers often desire  finer-grained control in literature search and
exploration tools \citep{Dunne2012ASE, neves2019evaluation, hope2020scisight}.
Context-dependent faceted search tools have the potential to fill this
gap. Finally, while we focus on literature search, faceted QBE also has the potential to 
serve in  other applications such as e-commerce search and product design 
\citep{hope2021scaling}.

The problem at the heart of faceted QBE is that of fine-grained document similarity. This
is of broader relevance to numerous other research problems. This work therefore 
promises to be of value in research problems such as that of multi-aspect 
reviewer assignments for papers \citep{Karimzadehgan2008aspectreview}, 
tracing the adoptions of ideas from the research literature \citep{cao2020idea}, 
or that of making causal inferences with matched scientific texts \citep[Sec: Applications and Simulations]{Roberts2020gendertextmat}. Appendix \ref{appendix-applications} elaborates on applications.

Despite this wide applicability of the problem, and a range of proposed
approaches, evaluation of these document retrieval/similarity methods remains a problem 
-- with prior work relying on weak sources of gold evaluation data or on expensive human evaluations.
A broad set of approaches evaluate systems against citations or combine
citation information with other incidental information such as section
headers as a means to determine citation intents acting as proxies for facets
\citep{ostendorff2020aspectcoling}.
While these sources may represent reasonable noisy signals \cite[Sec 4]{jurgens2018measuring} for model training, the noise in such approaches
limits their value as a method of system evaluation (Appenxix \ref{appendix-cite-eval} presents an analysis of citations as evaluation data). Other work has relied on extrinsic evaluations in a downstream ideation task \citep{chan2018solvent} or
based on feedback of users interacting with a recommendation system
\citep{ferosa2016tanmoy, abgaz2016evaluation}. However, such approaches require expensive evaluation protocols unsuitable for model development and system comparisons. Therefore, we believe our dataset will fill an important gap by presenting a pragmatic alternative to these extremes. 

Our publicly accessible test collection consists of 50 diverse research paper abstracts in English paired with a facet as
queries (\S\ref{sec-datasetdescr}).
The query facets  are chosen from one among
three facets from $\mathcal{F} = \{\texttt{background/objective}, \texttt{method},$
$\texttt{result}\}$, representing the three dominant aspects of methodological scientific 
research. Query abstracts are selected from the domains of machine learning and natural language processing.
Candidate pools of size $250$ ($8$ queries) or $100$ ($42$ queries)
are constructed from about $800,000$ computer science papers in the S2ORC
Corpus \citep{lo2020s2orc} using a diverse set of retrieval methods. 
Four graduate-level computer science annotators with research experience rated each 
candidate with respect to the query abstract and facet, with high agreement. Finally, we
also present a range of baseline experiments and analysis (\S\ref{sec-baseline-res}) 
indicating the phenomena that state of the art models fail to capture, presenting 
significant room for improvement.

\section{Task Formulations}
\label{sec-task-def}
We frame our task as one of retrieving scientific papers given a query paper
and additional information indicating the query facet.
In this work we operate at the level of abstracts rather than the body text of papers under the understanding that salient information about a paper is contained in the abstract, especially for the domains considered in this paper \cite{Medlar2019absbody}. Further, important applications in large deployed systems for paper recommendation and reviewer assignment operate at 
the level of paper abstracts indicating them to be an already useful choice \cite{S2Recommendation, ORExpertise}. Our decision is also motivated by the difficulty of annotating large
corpora at the body-text level, and mirrors a common choice in numerous prior 
work \citep{treccovid2021, Brown2019RELISH}. Note however, that our dataset 
is structured to leverage future full-text approaches
(See \S\ref{sec-datasetdescr}). Finally, \S\ref{sec-rel-analysis} presents a small-scale analysis of full-text vs abstract annotation consistency. In what follows, we also assume access to the title of the papers, although we drop it from the below discussion for brevity.

We denote a query
document with $Q$, a candidate document $C \in \mathcal{C}$. Every $Q$ or $C$,
consist of $N$ sentences $\langle S_1, S_2, \dots S_N\rangle$. We denote
\emph{facets}, $f$ from an inventory of labels $\mathcal{F}$, indicating the rhetorical elements of the document. We
denote a ranked list of the candidates for $Q$ and query facet
$f_q$ with $\langle r, C \rangle \in R_{Qf}$ where elements
denotes document $C$ at rank position $r$.
Now, we formulate two tasks that our test collection will effectively evaluate:

\begin{definition} Retrieval based on pre-defined facets:
Given query and candidate documents -- $Q$ and $\mathcal{C}$, with
sentences in both annotated with facet labels: $\langle(f_1, S_1), (f_2, S_2),
\dots(f_N, S_N)\rangle$ and a query facet $f_q$ a system must output the
ranking $R_{Qf}$.
\label{def-predefined-facets}
\end{definition}

\begin{definition} Retrieval by sentences:
Given query and candidate documents -- $Q$ and $\mathcal{C}$, and a subset of
sentences $\mathcal{S}_Q\subseteq Q$ based on which to
retrieve documents a system must output the ranking $R_Q$.
\label{def-query-by-sentence}
\end{definition} 

Def \ref{def-predefined-facets} corresponds most closely to faceted search as described  by \citet{weize2016facetedsearchthesis} and closest to work by
\citet{chan2018solvent}.  While Def \ref{def-query-by-sentence} represents a more general formulation not relying on pre-specified facet labels. Here, the sentences $\mathcal{S}_Q$ can be viewed as exemplar facet
sentences based on which results must be retrieved even while lacking any explicit facet specification.

One may view both Definition \ref{def-predefined-facets} and
\ref{def-query-by-sentence}
as instances of the QBE paradigm of retrieval
\citep{lissandrini2019exampletutorial}. One, at the level of
documents, using the document $Q$ as a query as in
\citet{el2011qbepapers} and \citet{zhu2014multiqbe}, and a second at the level of sentences denoting a facet of the paper. Broadly, we believe QBE to be
well suited to the problem of
faceted literature search given the difficulty of being able to specify in
keyword searches precisely the search intent and given that the meaning of
sentences denoting a facet are often dependent on the broader context of
the abstract. Further, we expect Definition \ref{def-query-by-sentence} to
have specific other advantages: users often tend to have different understanding of facets than
those defined by designers of the ranking system \citep{tunkelang2006dynamic} -- in our 
case we  expect that different sub-areas/areas of the literature will exemplify
different kinds of facets making it hard to pre-specify facets in a system.
Further, users often wish to explore the literature at different
levels of granularity than that possible with pre-defined facets
\cite[page 14]{hope2020scisight}, we expect QBE will allow users greater control to
select parts of an abstract expressing a facet at different levels of
granularity based on how they would like papers retrieved. Importantly however, note that while our dataset selects a specific set of facets for ease of annotation it facilitates evaluation and consequent model development of both task setups.

\section{Dataset Description}
\label{sec-datasetdescr}
In the construction of this test collection we relied on the Semantic Scholar
Open Research Corpus (S2ORC) \citep{lo2020s2orc} which provides a corpus of
$81.1$M English language research papers alongside a range of automatically
generated metadata including citation network information. We
choose to work with about
$800,000$ computer science papers in S2ORC
sourced from arXiv.\footnote{\url{https://arxiv.org/}} These papers were selected to ensure that the
full-body text of the papers was available, in addition to the abstract and
title, to facilitate potential future research.\footnote{\texttt{datasheet.md} in the dataset release documents detailed filtering steps used to obtain the ~800,000 documents. Our release also includes these ~800,000 documents.} Our queries were selected from domains of machine learning and NLP so that annotators would be familiar with the domain in question. 

\textbf{Facets: \label{sec-datasetdescr-facets}}
In this work a facet for a research paper corresponds to the dominant steps involved in carrying out scientific research --  the
identification of a research problem/question (\texttt{background/objective}), formation and testing of the hypothesis (\texttt{method}), and formation of conclusions (\texttt{results}). These facets are broadly defined as:
\begin{description}[noitemsep, nolistsep]
\item \texttt{background/objective}: Most often sets up the motivation for the
work, states how it relates to prior work and states the problem or
research question being asked. Henceforth, we refer to these as \texttt{background} facets.
\item \texttt{method}: Describes the method being proposed or used in the
paper. The method could be described at a very high level or it might be
specified at a very fine-grained level depending on the type of paper. Note
that our definition of methods is broad and will include methods of analysis
of a phenomena, a model, data, or procedural descriptions of the experiments
carried out. The specific interpretation of method also depends on the
type of paper (\S\ref{sec-datasetdescr-qsel}).
\item \texttt{result}: This may be a detailed statement of the findings of
analysis, a statement of results or a concluding set of statements based on the
type of paper.
\end{description}
Our corpus is labelled with facets predicted using the model of
\citet{cohan2019seqsentclf} into
the set of labels: \{\texttt{background}, \texttt{objective}, \texttt{method},
\texttt{result}, \texttt{other}\}. Incorrect facet labels for the query abstracts are manually corrected. Prior to relevance annotation,
\texttt{objective} and \texttt{background} are merged into one facet called
\texttt{background} as they were too similar to be distinguished for the purpose of document similarity. The \texttt{other} facet is not considered for annotation. These sentence facets were then provided as additional guidance during annotation, with query facet sentences being bolded to encourage attention to those parts of the document. 

\textbf{Query Abstract Selection: \label{sec-datasetdescr-qsel}}
We annotated a total of 50 query abstract-facet pairs from the ACL Anthology.\footnote{\url{https://www.aclweb.org/anthology/}} Of the 50, we annotated 16 abstracts with two different facets each (total of 32 query abstract-facet pairs), in order to
allow closer
analysis of the differences in retrieval performance for the same query abstract while varying the query facet (Figure \ref{fig-perq-specter}). The remaining
18 abstracts were annotated for a single query facet each. In total, our dataset
contains 16 \texttt{background} queries, 17 \texttt{method} queries, and
17 \texttt{result} queries; further statistics are provided in
Table \ref{tab:stats}.
Queries were selected to ensure coverage over a range of years (1995-2019) 
and to ensure a somewhat even distribution across query paper types, as coarsely divided into  ``resource/evaluation papers'', ``data-driven approach 
papers'' or ``theoretical papers''. This was ensured through randomly 
sampling a set of 100 articles over the time period and were manually 
filtered to ensure that each query corresponded to specific and non-trivial 
representations of multiple facets.  We include the query data distributions and the procedure for query selection in the 
annotator guidelines in our dataset release.\footnote{Annotator guideline and query metadata files in our dataset release: \texttt{ann\_guidelines.pdf} and
\texttt{queries-release.csv}}

\textbf{Candidate Pooling: \label{sec-datasetdescr-cpooling}}
Candidates per query are drawn from a corpus of about 800,000
computer science papers in the S2ORC corpus using the following pool of
methods:
TF-IDF, averaged \texttt{word2vec} embeddings, and TF-IDF weighted
\texttt{word2vec} based similarities run on titles, and abstracts giving us 
a set total of 6 methods. Further, a state-of-the-art \textsc{bert}  model, 
\textsc{specter} \citep{cohan2020specter},
trained for scientific paper representation using citation network signals 
was also part of the set of methods used to generate our pool.
Finally, papers cited in the query paper are also added to the pool, given
their likely relevance due to authors self selections. This set of
methods represents a diverse range of similarities with each of the methods
retrieving largely different candidate abstracts: The top-25 papers across
retrieval methods contained between 1-4 papers in common. 
For a set of 8 abstract-facet queries, we annotate pools of size 250.
These formed an initial exploratory set of annotations, and the remaining 42
queries were annotated with pools roughly of size 100. For queries with pools of size 250 we draw the top 33 papers from each retrieval method, similarly for queries with pools of 100 we draw the top 13 papers. The order in which we draw from the group of methods is
randomized for every query and in the case of a candidate already present
in our candidate pool we draw from further down the ranked list of a method.
Finally, while our task is framed as facet dependent, we use non-faceted
methods in the construction of our pool due to the lack of well-established faceted retrieval models. This choice also allowed us
to ensure an identical pool of candidates for being annotated with
respect to different facets -- providing for a richer evaluation setup. Statistics of the dataset are provided in Table \ref{tab:stats}.

\begin{table}[t]
\parbox{.45\linewidth}{
\centering
\captionsetup{justification=centering}
\caption{Statistics for the test collection.}
\scalebox{0.9}{
\begin{tabular}{r c r}
Statistic & & All \\\toprule
Query abstract-facet pairs & - &  50 \\ 
Unique query abstracts & - & 34 \\ 
Mean candidate pool size & - & 124.9 \\
Query-candidate pairs & - & 6244 \\\midrule
% & min & 93 \\
%Query abstract length in tokens & max & 241 \\
% & avg & 144.9 \\\hline
%& min & 26 \\
%Candidate abstract length in tokens & max & 590 \\
%& avg & 167.6 \\\hline
 & min & 12\\
Candidates rated +1  per query & max & 87 \\
 & avg & 36.9 \\\midrule
 & min & 1 \\
Candidates rated +2/+3 per query & max & 35 \\
 & avg & 9.8 
\end{tabular}}
\label{tab:stats}
}
\hfill
\parbox{.45\linewidth}{
\centering
\caption{Spearman's $\rho$, Krippendorff's $\alpha$, Cohen's $\kappa$, and \% agreement 
measures for relevances before and after the adjudication stage
of annotation. Given the ranking nature of the task, Spearman's $\rho$ presents the most apt measure of agreement.}
\scalebox{0.9}{
\begin{tabular}{rcccc}
           & \multicolumn{4}{c}{pre-adjudication}  \\
facet      & $\rho$ & $\alpha$ & $\kappa$ & \%     \\\toprule
background & 0.45     & 0.43 & 0.28 & 57.07           \\
method     & 0.31     & 0.26 & 0.20 & 69.60           \\
result     & 0.42     & 0.35 & 0.26 & 67.46           \\
           & \multicolumn{4}{c}{post-adjudication} \\\midrule
background & 0.73     & 0.72 &   0.62     & 77.68           \\
method     & 0.63     & 0.61 &   0.54     & 84.47           \\
result     & 0.70     & 0.67 &   0.59     & 83.53          
\end{tabular}}
\label{tab-agreement-measures}
}
\end{table}

\section{Dataset Annotation}
\textbf{Relevance Ratings: \label{sec-datasetdescr-rels}}
We choose to rate candidate documents on a graded scale from 0-3 with our
definitions for the scales depending on the facet. Broadly, we train 
annotators to rate structural/relational similarities between the candidate 
and query higher (3-2 ratings) than attribute/feature based similarities 
(1-0 ratings). This draws on motivations
from a range of literature highlighting the importance of structural similarities between ideas 
to creative activities like scientific research -- a focus of
this work \citep{douglas1990similarity, gentner2011companalogy, chan2018solvent,
Lavrac2020BisociativeLD}.  We illustrate the definitions with the \texttt{method} facet here:
\begin{description}[noitemsep, nolistsep]
    \item \texttt{Near Identical/+3}: +3 implies methods described share a
similar over-arching mechanistic similarity, further the methods must also 
be similar in terms of the details of the objects being manipulated.
    \item \texttt{Similar/+2}: +2 implies that the methods are mechanistically similar and the details are only comparable between the query and candidate.
    \item \texttt{Related/+1}: +1 is meant to encompass a wide range in being similar and can be hard to list at length. Common cases include: 1.\ Details of the two methods are similar but there is only high level mechanistic similarity. 2.\ Small or not-so-important mechanistic parts of the methods in two papers are similar. 3. Where query and candidate abstracts may vary in the level of granularity in which they describe a method and a high level similarity is the only one you can establish by reading the abstract.
\end{description}
Our relevance grades also include an \texttt{Unrelated/0} grade for documents deemed unrelated. We encourage readers to examine \texttt{ann\_guidelines.pdf} in our dataset release which details these further alongside examples for every case.

\textbf{Annotation Procedure: \label{sec-annotation}}
Our annotation was carried out by four graduate-level computer science annotators
(the lead authors, SM, TO, and 2 hired annotators) with experience
reading research papers in the selected domains.\footnote{Further details about annotators are included in the \texttt{datasheet.md} in the dataset release.} The annotation guidelines were developed over 4
iterations of repeated annotation and refinement of the guidelines. The hired
annotators were trained prior to annotation and demonstrated Spearman
Correlation based agreements of $0.5-0.7$ with an adjudicated training set of
examples. Annotators were paid an hourly wage of {USD} $22.5$ for a period of 3
weeks. All query-candidate pairs were annotated by two independent
annotators and a third adjudicator resolved the cases of difference between the
first two annotators. All annotations were carried out in Label
Studio\footnote{\url{https://labelstud.io/}} and annotators were only shown
paper titles and abstracts at all stages of annotation, hiding all other
metadata including authors, publication venues, and years.

\textbf{Relevance Analysis: \label{sec-rel-analysis}}
Given our two stage annotation process of gathering double annotation and an
adjudication stage, we report agreement metrics for both the stages. Table
\ref{tab-agreement-measures} presents these agreements.
Our pre-adjudication metrics are those between the two annotators involved in the
annotation. For the post-adjudication metrics, we report the mean metrics
between the adjudicated ratings and each of the two initial annotators ratings. 
We report 
Spearman rank-correlation coefficients, $\rho$, between annotators (pre-adjudication) and between annotators and adjudicated rankings (post-adjudication) produced by ratings. We believe $\rho$ to be the most apt measure of agreement given our ranking task, where we are most interested in establishing a relative similarity between papers. Our reporting also follows work in rating sentence similarities which reports correlations between annotators as a measure of agreement \cite{agirre2016semeval}. Additionally, we report Krippendorff's $\alpha$ with an ordinal distance function to measure agreement in absolute terms while taking into account an ordinal relevance scale, Cohens $\kappa$ to measure agreement while not taking into account the ordinal nature of relevance levels i.e.\ treating ratings as categorical labels, and simple percent agreements as an illustration of the fraction of data which needed adjudication. All of the
metrics in Table \ref{tab-agreement-measures} represent median values across
the per-query metrics. It was also permissible
for the adjudicator to entirely over-rule both annotators ratings for a candidate.
Across all facets, we saw this happen very rarely, 2-3\% on average per query. 

\begin{figure*}[tb!]
    \centering
\includegraphics[width=.3\textwidth]
{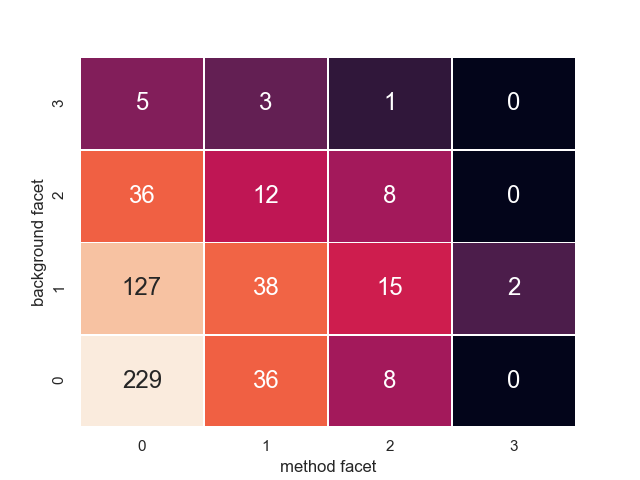}
\includegraphics[width=.3\textwidth]
{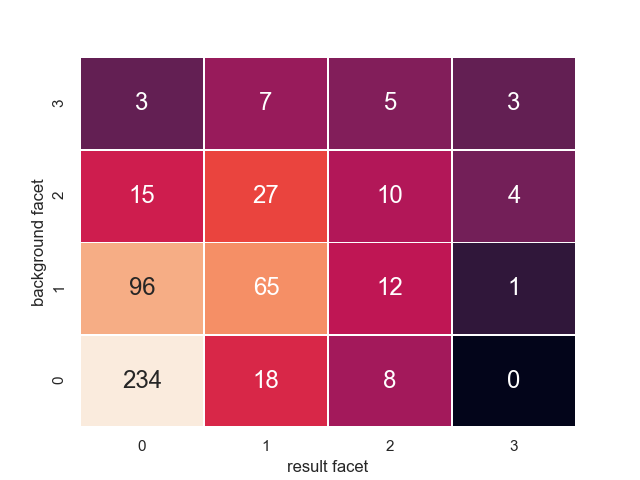}
\includegraphics[width=.3\textwidth]
{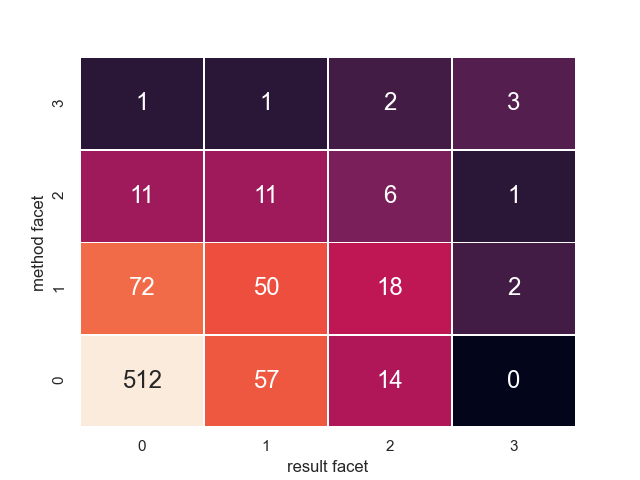}
\caption{Number of candidates labeled with
a particular facet relevance (scale of 0-3) for the documents labeled with two
facets. Note that while some between
facets correlation exists (along the diagonals), many candidates are relevant
only to one facet. Also note that this differs by facet pair.}
    \label{fig-double-corrs}
\end{figure*}

Based on the observed values of $\rho$, and instances of over-ruling in adjudication we believe that annotators are able to consistently establish a relative similarity between papers and indicate strong agreement with an adjudicated set of ratings. 
Based on qualitative observations in the annotation process we noted the primary case of disagreement between annotators. Disagreements occurred most where a single facet was representative of multiple
different finer-grained aspects. In these cases annotators initially focus on one of the
aspects in making their annotations, when made aware of other aspects during adjudication, annotators readily accepted a different judgement. Appendix \ref{sec-multi-aspect} presents an example. We believe this speaks to the effectiveness of an adjudication step. Finally, Figure \ref{fig-double-corrs}, indicates the set of relevances
for query abstracts annotated with two facets -- we see that while some candidates are
correlated in relevance others are only relevant to the query in one facet.

\textit{Full-text vs abstract annotations}: To examine the effect of annotating the abstract of papers instead of full-text of papers we also conduct a small scale study to examine the extent of differences between the relevances produced by them. This is done by a single expert annotator annotating relevance based on abstract and full-text separately for 9 query abstracts (3 from each facet) and 5 candidate abstracts each (45 pairs). We refer readers to Appendix \ref{appendix-ft-abs} for the details of the annotation setup. Agreement was measured between the relevances produced based on the abstract text and the full-text: Spearmans $\rho=0.78$, Krippendorff's $\alpha=0.77$, Cohens $\kappa=0.63$, and $\%$ agreement of $73\%$. Full-text annotation was performed at the rate of about 5 minutes/pair, on the other hand abstract ratings took 15 seconds/pair. While our metrics indicates a less than perfect match between the abstract ratings vs full-text these metrics indicate strong agreement between the two. We believe this presents a reasonable trade-off between expense and completeness of annotation. Further, prior work on a similar scientific similarity annotation task noted low-agreements between expert annotators in annotation of full-texts \cite[Sec 7]{wadden2020fact}, indicating that use of full-text does not necessarily translate into higher quality annotations.

\textbf{Comparison to Other Datasets: \label{sec-other-data}}
To the best of our knowledge, \citet{neves2019evaluation} and \citet{chan2018solvent} present the only other datasets of similar structure to the one we present:
\begin{description}[noitemsep, nolistsep]
    \item[Neves 2019:] \citet{neves2019evaluation} annotate a set of 8 query papers with 70-90 candidates per query from biomedicine for similarity of the ``papers research goal''.
    \item[SOLVENT:] \citet[Sec 4.1]{chan2018solvent} annotate a small scale dataset of 50 social computing papers for similarity along ``purpose'' or ``mechanism'' facets. Here, each of the 50 papers is annotated for relevance with respect to the other 49 papers in the set of 50 papers. 
\end{description}
Both these datasets intend at presenting small scale evaluations in the context of other work. Our work, in contrast, presents a substantially larger resource created over multiple annotation passes, annotates similarity across multiple facets, and presents a reusable dataset. Several other datasets bear partial resemblances to the work we present here, we discuss these in Section \ref{sec-related-work}.

\textbf{Annotation scalability:} We believe that a dataset like ours necessarily calls for expert annotation. Understanding the annotation guidelines alone required experience reading research papers. We believe these traits make it harder to scale the presented datasets using an un-trained crowd-sourcing based method. However, we believe certain aspects of our work can inform future work to scale datasets like ours. A dominant approach in building datasets involves annotation of instances by multiple annotators followed by a majority vote \cite{bowman2015large} or an average \cite{agirre2016semeval} to determine the ``gold'' label. Given the expense of creating repeated annotations when working with expert annotators we instead choose to use a adjudication step to create ``gold'' labels. We believe this helped us scale our approach considerably. If this step leveraged more experienced annotators to perform adjudication while lower experience annotators make initial judgements we believe it would be scalable beyond the approach presented here. This also follows prior work from \citet{nallapati2012skierarchy}, which leverages a hierarchy of annotator skills to scale a complex annotation task. Finally, we also note that large IR test collections call for competing systems to generate a pool of documents to judge \cite{treccovid2021}. We believe our dataset presents a robust evaluation set which may be leveraged for development of such methods.

\section{Experimental Results}
\label{sec-baseline-res}
Here, we establish baseline performances from a handful of standard and state of the art methods. 

\textbf{Baseline methods: \label{sec-baseline-descr}}
The methods we choose to evaluate capture a range of granularities and nature of methods: term based methods, pre-trained model based sentence representations, and whole abstract representation models. Appendix \ref{appendix-baseline-descr} describes each method in detail.
\begin{description}[noitemsep,nolistsep]
    \item Term-level baselines: \texttt{fabs\_tfidf}, \texttt{fabs\_bm25}, \texttt{fabs\_cbow200}, and \texttt{fabs\_tfidfcbow200} represent term-level baselines. These represent the document as sparse TF-IDF vectors, averaged bag of word representations, and weighted averaged bag of word representations respectively. Each of them represent the query document as the representation for the sentences corresponding to the query facet sentences in the query abstract. Candidates are represented by their whole abstract representations. 
    \item Sentence-level baselines:  Here encode all query facet sentences and all candidate abstract sentences individually with a sentence encoder, and then use the maximum pairwise sentence cosine similarity between the query and candidate sentences to rank candidates. \texttt{SentBERT-PP}, \texttt{SentBERT-NLI}, \texttt{UnSimCSE}, \texttt{SuSimCSE} (unsupervised and supervised \texttt{SimCSE}) represent state of the art sentence encoder baseline models \citep{reimers2019sentencebert, gao2021simcse}. Each of the models here represent models trained in a different manner or on different training sets.
    \item Abstract-level baselines: \textsc{specter}\ and \textsc{specter-id}\ represent whole abstract level representations \citep{cohan2020specter}. Both of these approaches represents a multi-layer transformer
based \textsc{SciBERT} model fine-tuned on citation network data. \textsc{specter} operates on titles and the whole abstract of
the papers. Both queries and candidates are represented by their \textsc{specter} embeddings. Note that \textsc{specter} was trained on a corpus of randomly selected scientific documents. We also re-implement and train a version of \textsc{specter} on about 660k computer science paper triples with identical hyper-parameters to \textsc{specter}, we call this in-domain model \textsc{specter-id}.
\end{description}
In re-ranking we use the L2 distance between the query and candidate vectors unless noted otherwise. Note 
here that while the term-level baselines are more similar to the the task formulated in Definition \ref{def-predefined-facets}, the sentence-level baselines solve the task in Definition \ref{def-query-by-sentence}. Further note that we make sure to include baseline methods from the above method types such that were not used for the constructions of pools. This is intended to evaluate the performance of methods which were not used for pool construction (\S\ref{sec-datasetdescr-cpooling}), thereby investigating the ability of the dataset to be re-used for evaluating future methods. These are represented by \texttt{fabs\_bm25}, \texttt{SentBERT} and \texttt{SimCSE} based methods, and the \textsc{specter-id} baseline. 
\begin{table*}[t]
\centering
\caption{Test set results for the set of baselines methods. Metrics (R-Precision, Precision and Recall at 20, \textsc{NDCG}$_{\%20}$) are computed based on a 2-fold cross-validation, represent averages over per-query metrics, and are reported as percentages. \textsc{specter-id} performance is reported over three training re-runs with underset standard-deviation, the remaining baselines are reported based on a single set of model parameters released by the respective authors.}
\scalebox{0.8}{
\begin{tabular}{rcccc|cccc}
                     & \multicolumn{4}{c}{\texttt{background}}   & \multicolumn{4}{c}{\texttt{method}}        \\ 
                     & RP     & P@20   & R@20   & NDCG$_{\%20}$   & RP     & P@20   & R@20  & NDCG$_{\%20}$ \\\toprule
\texttt{fabs\_tfidf}          & 23.35 & 27.19 & 45.80 &  57.97 & 09.30  & 09.83 & 34.75 & 31.20\\
\texttt{fabs\_bm25}           & 20.12 & 27.81 & 49.85 &  59.39 & 09.37 & 11.63 & 38.29 & 34.59\\
\texttt{fabs\_cbow200}        & 19.61 & 15.94 & 27.97 &  36.56 & 08.65 & 08.33 & 15.69 & 21.14\\
\texttt{fabs\_tfidfcbow200}   & 15.92 & 16.87 & 27.76 &  40.51 & 07.99 & 06.01 & 17.71 & 21.70\\
\texttt{SentBERT-PP}          & 21.24 & 28.75 & 46.67 &  60.80 & 10.00 & 10.83 & 36.30 & 33.40\\
\texttt{SentBERT-NLI}         & 19.02 & 25.00 & 40.13 &  54.23 & 09.11 & 11.46 & 02.89 & 31.10\\
\texttt{UnSimCSE-BERT}        & 18.15 & 23.44 & 36.05 &  51.59 & 08.86 & 09.65 & 27.92 & 31.23\\
\texttt{SuSimCSE-BERT}        & 19.22 & 22.81 & 46.75 &  55.22 & 08.58 & 09.76 & 29.01 & 30.88\\
\textsc{specter}              & \textbf{24.81} & \textbf{35.31} & \textbf{57.45} & 66.70 & {\bf 11.72} & {13.58} & {40.81} & 37.41\\
\textsc{specter-id}              & $\underset{\pm1.3}{24.55}$ & $\underset{\pm0.5}{34.17}$ & $\underset{\pm0.3}{53.26}$ &  $\underset{\pm1.71}{\bf 69.22}$ & $\underset{\pm0.3}{10.53}$ & $\underset{\pm1.21}{\bf 16.22}$ & $\underset{\pm3.6}{\bf 44.59}$ & $\underset{\pm0.78}{\bf 42.76}$\\
\multicolumn{1}{l}{} & \multicolumn{4}{c}{\texttt{result}}       & \multicolumn{4}{c}{\textit{Aggregated}}           \\ 
                     & RP     & P@20   & R@20   & NDCG$_{\%20}$  & RP     & P@20   & R@20     & NDCG$_{\%20}$\\\toprule
\texttt{fabs\_tfidf}          & 11.35 & 16.28 & 38.57  & 41.24 & 14.59 & 17.64 & 39.69 & 43.19\\
\texttt{fabs\_bm25}           & 11.31 & 20.00 & 40.40 & 45.07 & 13.50 & 19.69 & 42.73 & 46.06\\
\texttt{fabs\_cbow200}        & 11.16 & 10.42 & 23.44 & 30.93 & 13.08 & 11.47 & 22.23 & 29.36\\
\texttt{fabs\_tfidfcbow200}   & 10.43 & 10.69 & 24.39 & 32.79 & 11.38 & 11.09 & 23.13 & 31.42\\
\texttt{SentBERT-PP}          & 13.60 & 19.83 & 41.73 & 52.35 & 14.83 & 19.62 & 41.41 & 48.57\\
\texttt{SentBERT-NLI}         & 14.23 & 22.05 & 46.99 & 51.30 & 14.04 & 19.42 & 38.67 & 45.39\\
\texttt{UnSimCSE-BERT}        & 12.00 & 19.58 & 38.95 & 45.55 & 12.92 & 17.41 & 34.43 & 42.59\\
\texttt{SuSimCSE-BERT}        & 12.37 & 18.58 & 39.76 & 44.93 & 13.33 & 16.95 & 34.83 & 43.45\\
\textsc{specter}              & {18.62} & {23.78} & {52.72} & 56.67 & {18.29} & {23.97} & {50.14} & 53.28\\
\textsc{specter-id}             & $\underset{\pm0.92}{\bf 20.09}$ & $\underset{\pm0.45}{\bf 27.36}$ & $\underset{\pm3.04}{\bf 58.74}$ & $\underset{\pm1.31}{\bf 60.40}$ &  $\underset{\pm0.79}{\bf 18.32}$ & $\underset{\pm0.22}{\bf 25.74}$ & $\underset{\pm1.54}{\bf 52.12}$ & $\underset{\pm0.70}{\bf 57.22}$
\end{tabular}   
\label{tab-baseline-results}}
\end{table*}

\textbf{Re-ranking Results:}
Table \ref{tab-baseline-results} denotes performance on the test set for each facet 
independently and aggregated in the \textit{Aggregated} columns. In reporting results, we report R-Precision, Precision@20, recall@20, and \textsc{ndcg}@k. For \textsc{ndcg}@k, we follow \citet{wang13ndcg}, and choose $k=p*|\mathcal{C}|$ where $p\in(0,1)$. \textsc{ndcg}$_{\%20}$ therefore refers to \textsc{ndcg} computed at 20\% of the pool size for a query. This is apt since our queries don't have identical pool sizes. Appendix \ref{appendix-extended-results} presents an extended result table.

\textit{Overall results:} First, in line with the strong performance of
pre-trained language model representations for a range of tasks, we note
broadly the stronger performance of \textsc{specter-id} and \texttt{SentBERT} models  compared to term and static embedding based baselines. However, note that the sentence level \texttt{SentBERT} models underperform models which incorporate the whole abstract context. Also note that training on in-domain data allows \textsc{specter-id} some gains over \textsc{specter}. In examining facet dependent performance, we note the stronger performance of all the methods on the
\texttt{background} facet, which is expected due to the stronger correlation between background sentences and the general topic of the paper. Next, we note the
consistently poorer performance of all methods on the \texttt{method} facet, providing clear room for improvement. As
might be noted from our relevance rating guidelines
(\S\ref{sec-datasetdescr-rels}), we rate ``mechanistic'' notions of similarity
for the method facet. Given this relational nature of similarity, we expect
methods relying on whole paragraph or term level representations to perform
poorly on this facet. We note results midway between the other facets for \texttt{result} -- this is due to some \texttt{result}s being easy to be judged similar based on term overlaps while others require deeper a understanding of the query (See Appendix \ref{appendix-analysis}). Finally, we make special note of the poor performance of recent state of the art models \texttt{SimCSE}, and \textsc{specter} on overall performances, specially so in \texttt{method} and \texttt{result} facets -- we believe this offers future work substantial room for improvement. 

\begin{figure}[t]
     \centering
\includegraphics[width=0.5\textwidth]
{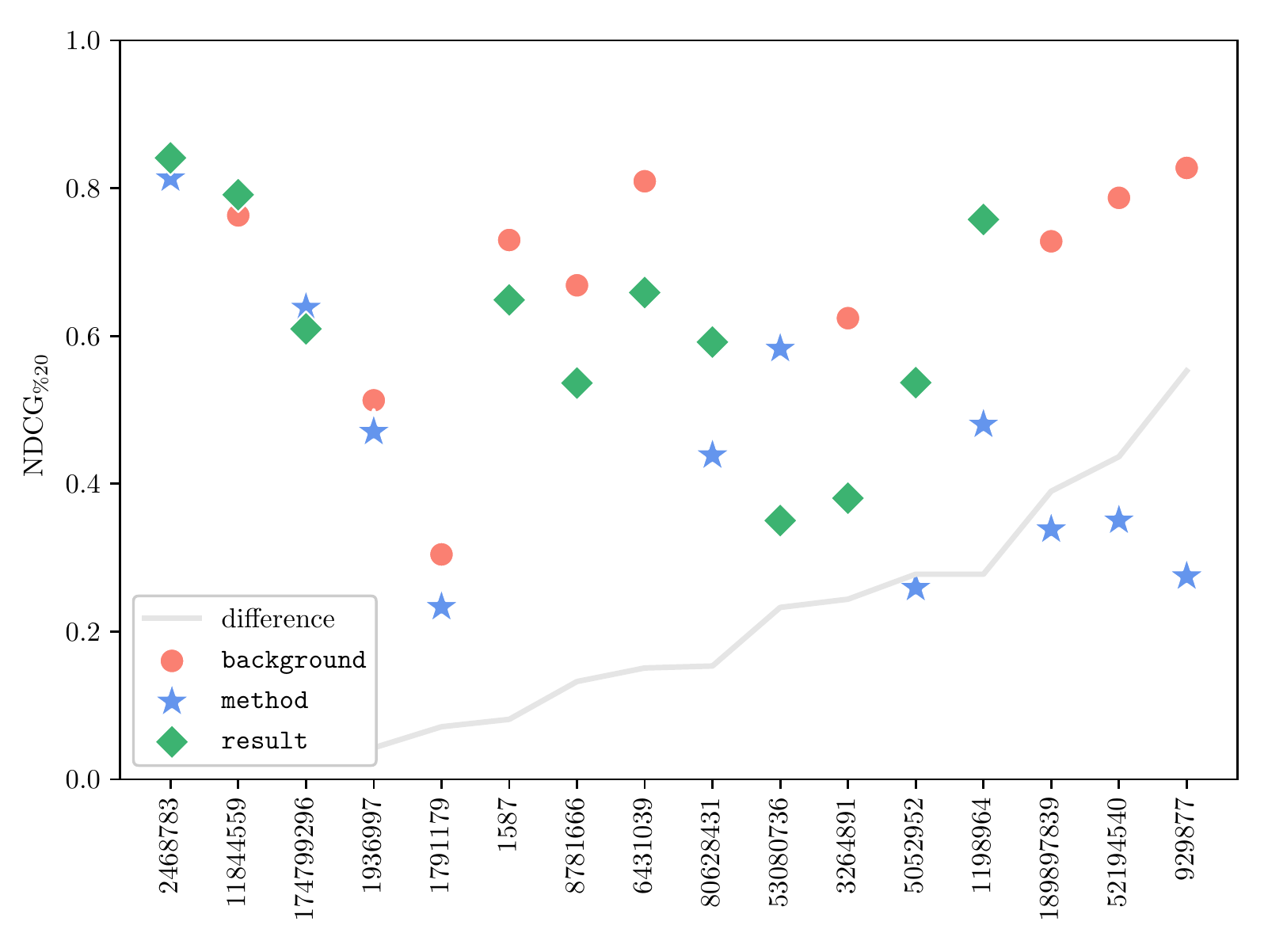}
     \caption{Per-query performance (NDCG$_{\%20}$) for \textsc{specter-id} on the set of papers which have been annotated with two facets (\S\ref{sec-datasetdescr}). Performances are means across 3 separate model training re-runs. The query papers are sorted by difference in performance on the facets.}
     \label{fig-perq-specter}
 \end{figure}

Since we annotate multiple queries per abstract we also present per-query
results for the best performing baseline, \textsc{specter-id}, on this set of queries in
Figure \ref{fig-perq-specter}. We note here the difference in performance by facet for \textsc{specter-id}, an un-faceted model. Here, performing well on one facet does not always lead to strong performance on other facets indicating room for improvement with models which incorporate finer-grained conditional similarities into their rankings.

\textit{Reusabilty:} Since we evaluate methods not used for pool construction (i.e \texttt{fabs\_bm25}, \texttt{SentBERT} and \texttt{SimCSE} methods, and \textsc{specter-id}), we examine their performance. Here, both \texttt{fabs\_bm25} and \textsc{specter-id} outperform corresponding methods of their method-type used for pool construction (i.e.\ \texttt{fabs\_tfidf} and \textsc{specter}). Further \texttt{SentBERT-PP}, a method representing a different class of method than those used for pool construction also outperforms the kinds of methods used for pool construction, notably \texttt{fabs\_tfidf}. We believe this indicates the lack of a serious bias of the dataset toward methods used for pool-construction, allowing re-use for evaluating future methods. 

\section{Error Analysis}
\label{sec-analysis}
Based on a qualitative examination of per-query ranking performance of \texttt{abs\_tfidf}, \texttt{SentBERT-PP} and \textsc{specter-id} we outline the factors which lead the baseline models to underperform. We believe the incorporation of modeling to handle these phenomena will lead to improved performance on the task. While we provide a summary of here, Appendix \ref{appendix-analysis} provides more extensive examples.

\begin{description}[noitemsep, nolistsep]
    \item \textit{Salient Aspects:} One source of error is the inability of models to identify the most salient aspects for similarity -- representations capturing ``what the paper is about?'', often expressed only in part of a larger set of facet sentences.
    \item \textit{Multiple Aspects:} Within a given facet, papers often expressed multiple finer grained aspects, models however often only retrieved based on a single aspect.
    \item \textit{Domain specific similarities}: A set of errors also arise from the inability of models to determine similarity between technical concepts. For example, an inability to judge ``stacking'', ``ensemble strategy'', and ``bagging'' as similar.
    \item \textit{Mechanistic similarities:} Nearly all methods perform poorly in the case of determining mechanistic similarity in \texttt{method} facets. This often relies on determining similarity across a sequence of actions. This problem also bears resemblance to the challenging setup of retrieval based on movie plots as in \citet{arguello2021tip}.
    \item \textit{Context dependence of facets:} Faceted similarities as labelled here often also show context dependence on other facets, especially for \texttt{result} queries. Given that one major guideline for result similarity in our dataset is ``the same finding or conclusion'', being able to determine context similarity is important.
    \item \textit{Qualitative result statements:} \texttt{result} queries which summarize qualitative findings as, opposed to reports of performance on a dataset, often perform poorer, often requiring broader context and often lacking in term overlaps which may otherwise easily indicate relevance.
\end{description}
Therefore, the range of challenging phenomena captured in our dataset allow evaluation for novel modeling approaches to document similarity which in-turn translate into progress on a range of important problems. Appendix \ref{appendix-potential-training} indicates potential sources of training data which may be leveraged to train fine-grained document similarity models to overcome some of the above challenges.

\section{Related Work}
\label{sec-related-work}
Work presented here ties to that of information retrieval and scientific literature search communities. 

\textit{Faceted Search}: IR has considered the task of faceted search, where facets
have often been treated as fixed attributes of metadata
\citep{Kong2016thesis}, in line with a QBE setup our work provides a more semantic interpretation of a
facet tied to the rhetorical structure of the document. Other
similar work in IR comes from those aiming to diversify
search results along the specific aspects/intents of an under-specified keyword query
\citep{Agrawal2009diversifying}; one might consider the ``aspects'' in this line
of work to our ``facets''. Others have also explored dynamic generation of
facet like attributes for queries  \citep{Kong2016precision}. Partly similar to our task, \citet{Upadhyay2020AspectKB}, allow specification of aspects along side ad-hoc queries.

\textit{Query-by-Example}: A range of literature has also considered the QBE formulation applied
to a variety of different kinds of data, from graphs, text, music and to
archival image search \citep{lissandrini2019exampletutorial,
adewoye2020qbeimages, kamper2019musicqbe}. Both \citet{sarwar2020queryevent}
and \citet{tabib2020interactive} frame retrieval from text corpora using event
or syntactic representations as QBE. The closest work in QBE applied to
research papers search comes from \citet{el2011qbepapers, jacovi2021scalable} which considers
multiple document queries for research paper recommendation and
\citet{zhu2014multiqbe}, who additionally propose considering topic variety within
multiple query documents. While \citet{el2011qbepapers} evaluates their approach with a user study, \citet{jacovi2021scalable} employ document key-phrases as a proxy for finer grained relevance.

\textit{Literature search and recommendation:} Other related work to the one presented here comes from a range of work
exploring literature search. \citet{ferosa2016tanmoy} trained faceted
paper retrieval based on citation contexts in a paper. \citet{jain2018disentangled} train models to learn disentangled abstract representations trained on aspect-labelled data for biomedical randomized control trial papers.
\citet{chan2018solvent} also presents closely related work, where they explore
the problem of recommending analogically similar scientific articles, and wherein they also include a small-scale evaluation dataset comparable to the one
presented here. In similar vein \citet{neves2019evaluation}, extensively
evaluate a range of methods intended to extract rhetorical structure elements
for faceted scientific paper search and evaluate it with a small scale dataset
of biomedical publications labelled for fine grained facet similarity.
\citet{hope2020scisight} allows exploratory search using multi-facet characterizations of
scientific articles for the COVID19 research literature.
Faceted document similarity for articles has been explored most recently in by \citet{ostendorff2020contextual,ostendorff2020aspectcoling} and \citet{kobayashi2018citation}. Both \citet{ostendorff2020aspectcoling} and \citet{kobayashi2018citation} evaluate using the task of predicting facets of similarity for papers cited in a particular section of a research article. \citet{kobayashi2018citation} further evaluate on a context-dependent co-citation ranking task as well. Work presented in \citet{ostendorff2020contextual} and \citet{kobayashi2018citation} present tasks most similar to ours while lacking in manually annotated datasets.

A range of work also explores the problem of document representations for an unfaceted content based recommendations in scientific documents \citep{cohan2020specter,bhagavatula2018content} and are often evaluated using citation or paper recommendation tasks ~
--
\citet{farber2020citation} provide an extensive survey of this literature. 

\textit{Other Datasets}: A handful of other work also bears resemblance to aspects of our dataset. \citet{Brown2019RELISH}, present a large expert annotated biomedical dataset intended to benchmark content based literature recommendation, however the dataset is annotated at the whole abstract level unlike faceted relevances as presented here. Datasets intended to match citation context sentences to the abstract sentences of the papers they cite share  some similarity with the task of querying with sentences of a facet \citep{Chandrasekaran2019Scisumm}. However they represent a somewhat simpler and different task, requiring to match sentences from full-text  to the citance. In similar vein, our task setup also resembles that of claim verification as proposed in \citet{wadden2020fact}. This setup however while serving the different goal of claim verifications also deals with atomic facts as opposed to more complex context dependent scientific paper facets as in our setup. Finally, datasets intended for citation intent prediction \citep{jurgens2018measuring,nanba1999towards, teufel2006automatic} mainly focus on a classification task involving predicting citation-intent given pairs of papers as opposed to retrieving papers while conditioning on a paper and a facet.

\section{Conclusions}
In this work we formalize the task of faceted Query by Example in the context of scientific literature search and highlight several important related problems our dataset can facilitate progress on. Given the problems of scientific search, the inability to articulate keyword queries, context dependence, and desire for exploratory search, we believe the faceted QBE formulation provides meaningful benefit. We provide a expert annotated test collection for the evaluation of the faceted QBE task. While prior work on the problem has often relied upon small evaluation sets, silver evaluations based on citations or keyword based relevance, or expensive human evaluation we believe our dataset provides a pragmatic alternative which will facilitate comparison and development of models. Finally we evaluate performance with a host of strong baseline approaches and highlight the challenging aspects of the dataset and the faceted QBE problem in general, and note that the dataset offers significant room for improvement.

\section{Acknowledgements}
We are grateful to the members of IESL at UMass Amherst and the Olivetti Group at MIT for helpful discussions in formulating this project. We also appreciate the contributions of our annotators in creating the annotation guidelines and the dataset. We also thank anonymous reviewers for their suggestions for making our work stronger. This work was supported in part by the National Science Foundation under Grant No. DMR-1922090, the Chan Zuckerberg Initiative under the project Scientific Knowledge Base Construction, the International Business Machines Corporation Cognitive Horizons Network under the project Knowledge Extraction, Representation, and Reasoning, and the Center for Intelligent Information Retrieval. Any opinions, findings and conclusions or recommendations expressed in this material are those of the authors and do not necessarily reflect those of the sponsors.

\bibliographystyle{ACM-Reference-Format}
\bibliography{faceted-ret-test-col}

%%% -*-BibTeX-*-
%%% Do NOT edit. File created by BibTeX with style
%%% ACM-Reference-Format-Journals [18-Jan-2012].

\begin{thebibliography}{70}

%%% ====================================================================
%%% NOTE TO THE USER: you can override these defaults by providing
%%% customized versions of any of these macros before the \bibliography
%%% command.  Each of them MUST provide its own final punctuation,
%%% except for \shownote{}, \showDOI{}, and \showURL{}.  The latter two
%%% do not use final punctuation, in order to avoid confusing it with
%%% the Web address.
%%%
%%% To suppress output of a particular field, define its macro to expand
%%% to an empty string, or better, \unskip, like this:
%%%
%%% \newcommand{\showDOI}[1]{\unskip}   % LaTeX syntax
%%%
%%% \def \showDOI #1{\unskip}           % plain TeX syntax
%%%
%%% ====================================================================

\ifx \showCODEN    \undefined \def \showCODEN     #1{\unskip}     \fi
\ifx \showDOI      \undefined \def \showDOI       #1{#1}\fi
\ifx \showISBNx    \undefined \def \showISBNx     #1{\unskip}     \fi
\ifx \showISBNxiii \undefined \def \showISBNxiii  #1{\unskip}     \fi
\ifx \showISSN     \undefined \def \showISSN      #1{\unskip}     \fi
\ifx \showLCCN     \undefined \def \showLCCN      #1{\unskip}     \fi
\ifx \shownote     \undefined \def \shownote      #1{#1}          \fi
\ifx \showarticletitle \undefined \def \showarticletitle #1{#1}   \fi
\ifx \showURL      \undefined \def \showURL       {\relax}        \fi
% The following commands are used for tagged output and should be
% invisible to TeX
\providecommand\bibfield[2]{#2}
\providecommand\bibinfo[2]{#2}
\providecommand\natexlab[1]{#1}
\providecommand\showeprint[2][]{arXiv:#2}

\bibitem[\protect\citeauthoryear{Abgaz, O'Donoghue, Smorodinnikov, and
  Hurley}{Abgaz et~al\mbox{.}}{2016}]%
        {abgaz2016evaluation}
\bibfield{author}{\bibinfo{person}{Yalemisew Abgaz}, \bibinfo{person}{Diarmuid
  O'Donoghue}, \bibinfo{person}{Dmitry Smorodinnikov}, {and}
  \bibinfo{person}{Donny Hurley}.} \bibinfo{year}{2016}\natexlab{}.
\newblock \showarticletitle{Evaluation often Analogical Inferences Formed from
  Automatically Generated Representations of Scientific Publications}.
\newblock  (\bibinfo{year}{2016}).
\newblock


\bibitem[\protect\citeauthoryear{Adewoye, Han, Ruest, Milligan, Fritz, and
  Lin}{Adewoye et~al\mbox{.}}{2020}]%
        {adewoye2020qbeimages}
\bibfield{author}{\bibinfo{person}{Tobi Adewoye}, \bibinfo{person}{Xiao Han},
  \bibinfo{person}{Nick Ruest}, \bibinfo{person}{Ian Milligan},
  \bibinfo{person}{Samantha Fritz}, {and} \bibinfo{person}{Jimmy Lin}.}
  \bibinfo{year}{2020}\natexlab{}.
\newblock \showarticletitle{Content-Based Exploration of Archival Images Using
  Neural Networks}. ACM/IEEE.
\newblock
\urldef\tempurl%
\url{https://dl.acm.org/doi/10.1145/3383583.3398577}
\showURL{%
\tempurl}


\bibitem[\protect\citeauthoryear{Agirre, Banea, Cer, Diab, Gonzalez~Agirre,
  Mihalcea, Rigau~Claramunt, and Wiebe}{Agirre et~al\mbox{.}}{2016}]%
        {agirre2016semeval}
\bibfield{author}{\bibinfo{person}{Eneko Agirre}, \bibinfo{person}{Carmen
  Banea}, \bibinfo{person}{Daniel Cer}, \bibinfo{person}{Mona Diab},
  \bibinfo{person}{Aitor Gonzalez~Agirre}, \bibinfo{person}{Rada Mihalcea},
  \bibinfo{person}{German Rigau~Claramunt}, {and} \bibinfo{person}{Janyce
  Wiebe}.} \bibinfo{year}{2016}\natexlab{}.
\newblock \showarticletitle{Semeval-2016 task 1: Semantic textual similarity,
  monolingual and cross-lingual evaluation}. In
  \bibinfo{booktitle}{\emph{SemEval-2016}}. ACL.
\newblock


\bibitem[\protect\citeauthoryear{Agrawal, Gollapudi, Halverson, and
  Ieong}{Agrawal et~al\mbox{.}}{2009}]%
        {Agrawal2009diversifying}
\bibfield{author}{\bibinfo{person}{Rakesh Agrawal}, \bibinfo{person}{Sreenivas
  Gollapudi}, \bibinfo{person}{Alan Halverson}, {and} \bibinfo{person}{Samuel
  Ieong}.} \bibinfo{year}{2009}\natexlab{}.
\newblock \showarticletitle{Diversifying Search Results}. In
  \bibinfo{booktitle}{\emph{Proceedings of the Second ACM International
  Conference on Web Search and Data Mining}} \emph{(\bibinfo{series}{WSDM
  '09})}. \bibinfo{publisher}{Association for Computing Machinery},
  \bibinfo{address}{New York, NY, USA}, \bibinfo{pages}{5–14}.
\newblock
\showISBNx{9781605583907}
\urldef\tempurl%
\url{https://doi.org/10.1145/1498759.1498766}
\showDOI{\tempurl}


\bibitem[\protect\citeauthoryear{Antonello, Beckage, Turek, and Huth}{Antonello
  et~al\mbox{.}}{2021}]%
        {antonello2021selecting}
\bibfield{author}{\bibinfo{person}{Richard Antonello}, \bibinfo{person}{Nicole
  Beckage}, \bibinfo{person}{Javier Turek}, {and} \bibinfo{person}{Alexander
  Huth}.} \bibinfo{year}{2021}\natexlab{}.
\newblock \showarticletitle{Selecting Informative Contexts Improves Language
  Model Fine-tuning}. In \bibinfo{booktitle}{\emph{ACL}}.
  \bibinfo{publisher}{Association for Computational Linguistics},
  \bibinfo{address}{Online}.
\newblock
\urldef\tempurl%
\url{https://aclanthology.org/2021.acl-long.87}
\showURL{%
\tempurl}


\bibitem[\protect\citeauthoryear{Arguello, Ferguson, Fine, Mitra, Zamani, and
  Diaz}{Arguello et~al\mbox{.}}{2021}]%
        {arguello2021tip}
\bibfield{author}{\bibinfo{person}{Jaime Arguello}, \bibinfo{person}{Adam
  Ferguson}, \bibinfo{person}{Emery Fine}, \bibinfo{person}{Bhaskar Mitra},
  \bibinfo{person}{Hamed Zamani}, {and} \bibinfo{person}{Fernando Diaz}.}
  \bibinfo{year}{2021}\natexlab{}.
\newblock \showarticletitle{Tip of the Tongue Known-Item Retrieval: A Case
  Study in Movie Identification}. In \bibinfo{booktitle}{\emph{Proceedings of
  the 6th international ACM SIGIR Conference on Human Information Interaction
  and Retrieval}}. \bibinfo{publisher}{ACM}.
\newblock
\urldef\tempurl%
\url{https://dlnext.acm.org/doi/10.1145/3406522.3446021}
\showURL{%
\tempurl}


\bibitem[\protect\citeauthoryear{Berger, Zavrel, and Groth}{Berger
  et~al\mbox{.}}{2020}]%
        {berger2020effective}
\bibfield{author}{\bibinfo{person}{Mark Berger}, \bibinfo{person}{Jakub
  Zavrel}, {and} \bibinfo{person}{Paul Groth}.}
  \bibinfo{year}{2020}\natexlab{}.
\newblock \showarticletitle{Effective distributed representations for academic
  expert search}. In \bibinfo{booktitle}{\emph{Proceedings of the First
  Workshop on Scholarly Document Processing}}. \bibinfo{publisher}{Association
  for Computational Linguistics}, \bibinfo{address}{Online},
  \bibinfo{pages}{56--71}.
\newblock
\urldef\tempurl%
\url{https://doi.org/10.18653/v1/2020.sdp-1.7}
\showDOI{\tempurl}


\bibitem[\protect\citeauthoryear{Bhagavatula, Feldman, Power, and
  Ammar}{Bhagavatula et~al\mbox{.}}{2018}]%
        {bhagavatula2018content}
\bibfield{author}{\bibinfo{person}{Chandra Bhagavatula},
  \bibinfo{person}{Sergey Feldman}, \bibinfo{person}{Russell Power}, {and}
  \bibinfo{person}{Waleed Ammar}.} \bibinfo{year}{2018}\natexlab{}.
\newblock \showarticletitle{Content-Based Citation Recommendation}. In
  \bibinfo{booktitle}{\emph{Proceedings of the 2018 Conference of the North
  {A}merican Chapter of the Association for Computational Linguistics: Human
  Language Technologies, Volume 1 (Long Papers)}}.
  \bibinfo{publisher}{Association for Computational Linguistics},
  \bibinfo{address}{New Orleans, Louisiana}, \bibinfo{pages}{238--251}.
\newblock
\urldef\tempurl%
\url{https://doi.org/10.18653/v1/N18-1022}
\showDOI{\tempurl}


\bibitem[\protect\citeauthoryear{Bowman, Angeli, Potts, and Manning}{Bowman
  et~al\mbox{.}}{2015}]%
        {bowman2015large}
\bibfield{author}{\bibinfo{person}{Samuel~R. Bowman}, \bibinfo{person}{Gabor
  Angeli}, \bibinfo{person}{Christopher Potts}, {and}
  \bibinfo{person}{Christopher~D. Manning}.} \bibinfo{year}{2015}\natexlab{}.
\newblock \showarticletitle{A large annotated corpus for learning natural
  language inference}. In \bibinfo{booktitle}{\emph{Proceedings of the 2015
  Conference on Empirical Methods in Natural Language Processing}}.
  \bibinfo{publisher}{Association for Computational Linguistics},
  \bibinfo{address}{Lisbon, Portugal}, \bibinfo{pages}{632--642}.
\newblock
\urldef\tempurl%
\url{https://doi.org/10.18653/v1/D15-1075}
\showDOI{\tempurl}


\bibitem[\protect\citeauthoryear{Brown, Consortium, and Zhou}{Brown
  et~al\mbox{.}}{2019}]%
        {Brown2019RELISH}
\bibfield{author}{\bibinfo{person}{Peter Brown}, \bibinfo{person}{RELISH
  Consortium}, {and} \bibinfo{person}{Yaoqi Zhou}.}
  \bibinfo{year}{2019}\natexlab{}.
\newblock \showarticletitle{{Large expert-curated database for benchmarking
  document similarity detection in biomedical literature search}}.
\newblock \bibinfo{journal}{\emph{Database}}  \bibinfo{volume}{2019}
  (\bibinfo{date}{10} \bibinfo{year}{2019}).
\newblock
\showISSN{1758-0463}
\urldef\tempurl%
\url{https://doi.org/10.1093/database/baz085}
\showURL{%
\tempurl}


\bibitem[\protect\citeauthoryear{Cao, Cheng, Cen, McFarland, and Ren}{Cao
  et~al\mbox{.}}{2020}]%
        {cao2020idea}
\bibfield{author}{\bibinfo{person}{Hancheng Cao}, \bibinfo{person}{Mengjie
  Cheng}, \bibinfo{person}{Zhepeng Cen}, \bibinfo{person}{Daniel McFarland},
  {and} \bibinfo{person}{Xiang Ren}.} \bibinfo{year}{2020}\natexlab{}.
\newblock \showarticletitle{Will This Idea Spread Beyond Academia?
  Understanding Knowledge Transfer of Scientific Concepts across Text Corpora}.
  In \bibinfo{booktitle}{\emph{Findings of the Association for Computational
  Linguistics: EMNLP 2020}}. \bibinfo{publisher}{Association for Computational
  Linguistics}, \bibinfo{address}{Online}, \bibinfo{pages}{1746--1757}.
\newblock
\urldef\tempurl%
\url{https://doi.org/10.18653/v1/2020.findings-emnlp.158}
\showDOI{\tempurl}


\bibitem[\protect\citeauthoryear{Chakraborty, Krishna, Singh, Ganguly, Goyal,
  and Mukherjee}{Chakraborty et~al\mbox{.}}{2016}]%
        {ferosa2016tanmoy}
\bibfield{author}{\bibinfo{person}{Tanmoy Chakraborty}, \bibinfo{person}{Amrith
  Krishna}, \bibinfo{person}{Mayank Singh}, \bibinfo{person}{Niloy Ganguly},
  \bibinfo{person}{Pawan Goyal}, {and} \bibinfo{person}{Animesh Mukherjee}.}
  \bibinfo{year}{2016}\natexlab{}.
\newblock \showarticletitle{FeRoSA: A Faceted Recommendation System for
  Scientific Articles}. In \bibinfo{booktitle}{\emph{Proceedings, Part II, of
  the 20th Pacific-Asia Conference on Advances in Knowledge Discovery and Data
  Mining - Volume 9652}} \emph{(\bibinfo{series}{PAKDD 2016})}.
  \bibinfo{publisher}{Springer-Verlag}, \bibinfo{address}{Berlin, Heidelberg},
  \bibinfo{pages}{528–541}.
\newblock
\showISBNx{9783319317496}
\urldef\tempurl%
\url{https://doi.org/10.1007/978-3-319-31750-2_42}
\showDOI{\tempurl}


\bibitem[\protect\citeauthoryear{Chan, Chang, Hope, Shahaf, and Kittur}{Chan
  et~al\mbox{.}}{2018}]%
        {chan2018solvent}
\bibfield{author}{\bibinfo{person}{Joel Chan}, \bibinfo{person}{Joseph~Chee
  Chang}, \bibinfo{person}{Tom Hope}, \bibinfo{person}{Dafna Shahaf}, {and}
  \bibinfo{person}{Aniket Kittur}.} \bibinfo{year}{2018}\natexlab{}.
\newblock \showarticletitle{SOLVENT: A Mixed Initiative System for Finding
  Analogies between Research Papers}.
\newblock \bibinfo{journal}{\emph{Proc. ACM Hum.-Comput. Interact.}}
  \bibinfo{volume}{2}, \bibinfo{number}{CSCW}, Article \bibinfo{articleno}{31}
  (\bibinfo{date}{Nov.} \bibinfo{year}{2018}), \bibinfo{numpages}{21}~pages.
\newblock
\urldef\tempurl%
\url{https://doi.org/10.1145/3274300}
\showDOI{\tempurl}


\bibitem[\protect\citeauthoryear{Chandrasekaran, Yasunaga, Radev, Freitag, and
  Kan}{Chandrasekaran et~al\mbox{.}}{2019}]%
        {Chandrasekaran2019Scisumm}
\bibfield{author}{\bibinfo{person}{Muthu~Kumar Chandrasekaran},
  \bibinfo{person}{Michihiro Yasunaga}, \bibinfo{person}{Dragomir Radev},
  \bibinfo{person}{Dayne Freitag}, {and} \bibinfo{person}{Min-Yen Kan}.}
  \bibinfo{year}{2019}\natexlab{}.
\newblock \showarticletitle{Overview and Results: CL-SciSumm Shared Task 2019}.
  In \bibinfo{booktitle}{\emph{In Proceedings of Joint Workshop on
  Bibliometric-enhanced Information Retrieval and NLP for Digital Libraries
  (BIRNDL 2019)}}.
\newblock
\urldef\tempurl%
\url{http://ceur-ws.org/Vol-2414/paper17.pdf}
\showURL{%
\tempurl}


\bibitem[\protect\citeauthoryear{Cohan, Beltagy, King, Dalvi, and Weld}{Cohan
  et~al\mbox{.}}{2019}]%
        {cohan2019seqsentclf}
\bibfield{author}{\bibinfo{person}{Arman Cohan}, \bibinfo{person}{Iz Beltagy},
  \bibinfo{person}{Daniel King}, \bibinfo{person}{Bhavana Dalvi}, {and}
  \bibinfo{person}{Dan Weld}.} \bibinfo{year}{2019}\natexlab{}.
\newblock \showarticletitle{Pretrained Language Models for Sequential Sentence
  Classification}. In \bibinfo{booktitle}{\emph{Proceedings of the 2019
  Conference on Empirical Methods in Natural Language Processing and the 9th
  International Joint Conference on Natural Language Processing
  (EMNLP-IJCNLP)}}. \bibinfo{publisher}{Association for Computational
  Linguistics}, \bibinfo{address}{Hong Kong, China},
  \bibinfo{pages}{3693--3699}.
\newblock
\urldef\tempurl%
\url{https://doi.org/10.18653/v1/D19-1383}
\showDOI{\tempurl}


\bibitem[\protect\citeauthoryear{Cohan, Feldman, Beltagy, Downey, and
  Weld}{Cohan et~al\mbox{.}}{2020}]%
        {cohan2020specter}
\bibfield{author}{\bibinfo{person}{Arman Cohan}, \bibinfo{person}{Sergey
  Feldman}, \bibinfo{person}{Iz Beltagy}, \bibinfo{person}{Doug Downey}, {and}
  \bibinfo{person}{Daniel Weld}.} \bibinfo{year}{2020}\natexlab{}.
\newblock \showarticletitle{{SPECTER}: Document-level Representation Learning
  using Citation-informed Transformers}. In
  \bibinfo{booktitle}{\emph{Proceedings of the 58th Annual Meeting of the
  Association for Computational Linguistics}}. \bibinfo{publisher}{Association
  for Computational Linguistics}, \bibinfo{address}{Online},
  \bibinfo{pages}{2270--2282}.
\newblock
\urldef\tempurl%
\url{https://doi.org/10.18653/v1/2020.acl-main.207}
\showDOI{\tempurl}


\bibitem[\protect\citeauthoryear{Dimitriadou, Papaemmanouil, and
  Diao}{Dimitriadou et~al\mbox{.}}{2014}]%
        {Dimitriadou2014EBE}
\bibfield{author}{\bibinfo{person}{Kyriaki Dimitriadou}, \bibinfo{person}{Olga
  Papaemmanouil}, {and} \bibinfo{person}{Yanlei Diao}.}
  \bibinfo{year}{2014}\natexlab{}.
\newblock \showarticletitle{Explore-by-Example: An Automatic Query Steering
  Framework for Interactive Data Exploration}. In
  \bibinfo{booktitle}{\emph{Proceedings of the 2014 ACM SIGMOD International
  Conference on Management of Data}} \emph{(\bibinfo{series}{SIGMOD '14})}.
  \bibinfo{publisher}{Association for Computing Machinery},
  \bibinfo{address}{New York, NY, USA}, \bibinfo{pages}{517–528}.
\newblock
\showISBNx{9781450323765}
\urldef\tempurl%
\url{https://doi.org/10.1145/2588555.2610523}
\showDOI{\tempurl}


\bibitem[\protect\citeauthoryear{Dunne, Shneiderman, Gove, Klavans, and
  Dorr}{Dunne et~al\mbox{.}}{2012}]%
        {Dunne2012ASE}
\bibfield{author}{\bibinfo{person}{Cody Dunne}, \bibinfo{person}{Ben
  Shneiderman}, \bibinfo{person}{Robert Gove}, \bibinfo{person}{Judith
  Klavans}, {and} \bibinfo{person}{Bonnie Dorr}.}
  \bibinfo{year}{2012}\natexlab{}.
\newblock \showarticletitle{Rapid understanding of scientific paper
  collections: Integrating statistics, text analytics, and visualization}.
\newblock \bibinfo{journal}{\emph{Journal of the American Society for
  Information Science and Technology}} \bibinfo{volume}{63},
  \bibinfo{number}{12} (\bibinfo{year}{2012}), \bibinfo{pages}{2351--2369}.
\newblock
\urldef\tempurl%
\url{https://doi.org/10.1002/asi.22652}
\showDOI{\tempurl}
\showeprint{https://onlinelibrary.wiley.com/doi/pdf/10.1002/asi.22652}


\bibitem[\protect\citeauthoryear{El-Arini and Guestrin}{El-Arini and
  Guestrin}{2011}]%
        {el2011qbepapers}
\bibfield{author}{\bibinfo{person}{Khalid El-Arini} {and}
  \bibinfo{person}{Carlos Guestrin}.} \bibinfo{year}{2011}\natexlab{}.
\newblock \showarticletitle{Beyond Keyword Search: Discovering Relevant
  Scientific Literature}. In \bibinfo{booktitle}{\emph{Proceedings of the 17th
  ACM SIGKDD International Conference on Knowledge Discovery and Data Mining}}
  \emph{(\bibinfo{series}{KDD '11})}. \bibinfo{publisher}{Association for
  Computing Machinery}, \bibinfo{address}{New York, NY, USA},
  \bibinfo{pages}{439–447}.
\newblock
\showISBNx{9781450308137}
\urldef\tempurl%
\url{https://doi.org/10.1145/2020408.2020479}
\showDOI{\tempurl}


\bibitem[\protect\citeauthoryear{Fadaee, Gureenkova, Rejon~Barrera, Schnober,
  Weerkamp, and Zavrel}{Fadaee et~al\mbox{.}}{2020}]%
        {fadaee2020new}
\bibfield{author}{\bibinfo{person}{Marzieh Fadaee}, \bibinfo{person}{Olga
  Gureenkova}, \bibinfo{person}{Fernando Rejon~Barrera},
  \bibinfo{person}{Carsten Schnober}, \bibinfo{person}{Wouter Weerkamp}, {and}
  \bibinfo{person}{Jakub Zavrel}.} \bibinfo{year}{2020}\natexlab{}.
\newblock \showarticletitle{A New Neural Search and Insights Platform for
  Navigating and Organizing {AI} Research}. In
  \bibinfo{booktitle}{\emph{Proceedings of the First Workshop on Scholarly
  Document Processing}}. \bibinfo{publisher}{Association for Computational
  Linguistics}, \bibinfo{address}{Online}, \bibinfo{pages}{207--213}.
\newblock
\urldef\tempurl%
\url{https://doi.org/10.18653/v1/2020.sdp-1.23}
\showDOI{\tempurl}


\bibitem[\protect\citeauthoryear{F{\"a}rber and Jatowt}{F{\"a}rber and
  Jatowt}{2020}]%
        {farber2020citation}
\bibfield{author}{\bibinfo{person}{Michael F{\"a}rber} {and}
  \bibinfo{person}{Adam Jatowt}.} \bibinfo{year}{2020}\natexlab{}.
\newblock \showarticletitle{Citation recommendation: approaches and datasets}.
\newblock \bibinfo{journal}{\emph{International Journal on Digital Libraries}}
  \bibinfo{volume}{21}, \bibinfo{number}{4} (\bibinfo{year}{2020}),
  \bibinfo{pages}{375--405}.
\newblock
\urldef\tempurl%
\url{https://doi.org/10.1007/s00799-020-00288-2}
\showDOI{\tempurl}


\bibitem[\protect\citeauthoryear{Gao, Yao, and Chen}{Gao et~al\mbox{.}}{2021}]%
        {gao2021simcse}
\bibfield{author}{\bibinfo{person}{Tianyu Gao}, \bibinfo{person}{Xingcheng
  Yao}, {and} \bibinfo{person}{Danqi Chen}.} \bibinfo{year}{2021}\natexlab{}.
\newblock \showarticletitle{{S}im{CSE}: Simple Contrastive Learning of Sentence
  Embeddings}. In \bibinfo{booktitle}{\emph{Proceedings of the 2021 Conference
  on Empirical Methods in Natural Language Processing}}.
  \bibinfo{publisher}{Association for Computational Linguistics},
  \bibinfo{address}{Online and Punta Cana, Dominican Republic},
  \bibinfo{pages}{6894--6910}.
\newblock
\urldef\tempurl%
\url{https://aclanthology.org/2021.emnlp-main.552}
\showURL{%
\tempurl}


\bibitem[\protect\citeauthoryear{Gentner and Forbus}{Gentner and
  Forbus}{2011}]%
        {gentner2011companalogy}
\bibfield{author}{\bibinfo{person}{Dedre Gentner} {and}
  \bibinfo{person}{Kenneth~D. Forbus}.} \bibinfo{year}{2011}\natexlab{}.
\newblock \showarticletitle{Computational models of analogy}.
\newblock \bibinfo{journal}{\emph{WIREs Cognitive Science}}
  \bibinfo{volume}{2}, \bibinfo{number}{3} (\bibinfo{year}{2011}),
  \bibinfo{pages}{266--276}.
\newblock
\urldef\tempurl%
\url{https://doi.org/10.1002/wcs.105}
\showDOI{\tempurl}
\showeprint{https://onlinelibrary.wiley.com/doi/pdf/10.1002/wcs.105}


\bibitem[\protect\citeauthoryear{Gil, Greaves, Hendler, and Hirsh}{Gil
  et~al\mbox{.}}{2014}]%
        {GilScienceReview}
\bibfield{author}{\bibinfo{person}{Yolanda Gil}, \bibinfo{person}{Mark
  Greaves}, \bibinfo{person}{James Hendler}, {and} \bibinfo{person}{Haym
  Hirsh}.} \bibinfo{year}{2014}\natexlab{}.
\newblock \showarticletitle{Amplify scientific discovery with artificial
  intelligence}.
\newblock \bibinfo{journal}{\emph{Science}} \bibinfo{volume}{346},
  \bibinfo{number}{6206} (\bibinfo{year}{2014}), \bibinfo{pages}{171--172}.
\newblock
\showISSN{0036-8075}
\urldef\tempurl%
\url{https://doi.org/10.1126/science.1259439}
\showDOI{\tempurl}
\showeprint{https://science.sciencemag.org/content/346/6206/171.full.pdf}


\bibitem[\protect\citeauthoryear{Hain, Jurowetzki, Buchmann, and Wolf}{Hain
  et~al\mbox{.}}{2020}]%
        {Hain2020TextbasedTS}
\bibfield{author}{\bibinfo{person}{Daniel~S. Hain}, \bibinfo{person}{Roman
  Jurowetzki}, \bibinfo{person}{Tobias Buchmann}, {and}
  \bibinfo{person}{Patrick Wolf}.} \bibinfo{year}{2020}\natexlab{}.
\newblock \showarticletitle{Text-based Technological Signatures and
  Similarities: How to create them and what to do with them}.
\newblock \bibinfo{journal}{\emph{ArXiv}}  \bibinfo{volume}{abs/2003.12303}
  (\bibinfo{year}{2020}).
\newblock


\bibitem[\protect\citeauthoryear{He, Wang, Zhang, Huang, and Caverlee}{He
  et~al\mbox{.}}{2020}]%
        {he2020parade}
\bibfield{author}{\bibinfo{person}{Yun He}, \bibinfo{person}{Zhuoer Wang},
  \bibinfo{person}{Yin Zhang}, \bibinfo{person}{Ruihong Huang}, {and}
  \bibinfo{person}{James Caverlee}.} \bibinfo{year}{2020}\natexlab{}.
\newblock \showarticletitle{{PARADE}: {A} {N}ew {D}ataset for {P}araphrase
  {I}dentification {R}equiring {C}omputer {S}cience {D}omain {K}nowledge}. In
  \bibinfo{booktitle}{\emph{Proceedings of the 2020 Conference on Empirical
  Methods in Natural Language Processing (EMNLP)}}.
  \bibinfo{publisher}{Association for Computational Linguistics},
  \bibinfo{address}{Online}, \bibinfo{pages}{7572--7582}.
\newblock
\urldef\tempurl%
\url{https://doi.org/10.18653/v1/2020.emnlp-main.611}
\showDOI{\tempurl}


\bibitem[\protect\citeauthoryear{Himanen, Geurts, Foster, and Rinke}{Himanen
  et~al\mbox{.}}{2019}]%
        {Himanen2019MSChallenges}
\bibfield{author}{\bibinfo{person}{Lauri Himanen}, \bibinfo{person}{Amber
  Geurts}, \bibinfo{person}{Adam~Stuart Foster}, {and} \bibinfo{person}{Patrick
  Rinke}.} \bibinfo{year}{2019}\natexlab{}.
\newblock \showarticletitle{Data-Driven Materials Science: Status, Challenges,
  and Perspectives}.
\newblock \bibinfo{journal}{\emph{Advanced Science}} \bibinfo{volume}{6},
  \bibinfo{number}{21} (\bibinfo{year}{2019}), \bibinfo{pages}{1900808}.
\newblock
\urldef\tempurl%
\url{https://doi.org/10.1002/advs.201900808}
\showDOI{\tempurl}
\showeprint{https://onlinelibrary.wiley.com/doi/pdf/10.1002/advs.201900808}


\bibitem[\protect\citeauthoryear{Hope, Portenoy, Vasan, Borchardt, Horvitz,
  Weld, Hearst, and West}{Hope et~al\mbox{.}}{2020}]%
        {hope2020scisight}
\bibfield{author}{\bibinfo{person}{Tom Hope}, \bibinfo{person}{Jason Portenoy},
  \bibinfo{person}{Kishore Vasan}, \bibinfo{person}{Jonathan Borchardt},
  \bibinfo{person}{Eric Horvitz}, \bibinfo{person}{Daniel~S Weld},
  \bibinfo{person}{Marti~A Hearst}, {and} \bibinfo{person}{Jevin West}.}
  \bibinfo{year}{2020}\natexlab{}.
\newblock \showarticletitle{SciSight: Combining faceted navigation and research
  group detection for COVID-19 exploratory scientific search}.
\newblock \bibinfo{journal}{\emph{arXiv preprint arXiv:2005.12668}}
  (\bibinfo{year}{2020}).
\newblock


\bibitem[\protect\citeauthoryear{Hope, Tamari, Kang, Hershcovich, Chan, Kittur,
  and Shahaf}{Hope et~al\mbox{.}}{2021}]%
        {hope2021scaling}
\bibfield{author}{\bibinfo{person}{Tom Hope}, \bibinfo{person}{Ronen Tamari},
  \bibinfo{person}{Hyeonsu Kang}, \bibinfo{person}{Daniel Hershcovich},
  \bibinfo{person}{Joel Chan}, \bibinfo{person}{Aniket Kittur}, {and}
  \bibinfo{person}{Dafna Shahaf}.} \bibinfo{year}{2021}\natexlab{}.
\newblock \bibinfo{title}{Scaling Creative Inspiration with Fine-Grained
  Functional Facets of Product Ideas}.
\newblock
\newblock
\showeprint[arxiv]{2102.09761}


\bibitem[\protect\citeauthoryear{Huang, Casey, G\l{}owacka, and Medlar}{Huang
  et~al\mbox{.}}{2019}]%
        {Medlar2019absbody}
\bibfield{author}{\bibinfo{person}{Chien-yu Huang}, \bibinfo{person}{Arlene
  Casey}, \bibinfo{person}{Dorota G\l{}owacka}, {and} \bibinfo{person}{Alan
  Medlar}.} \bibinfo{year}{2019}\natexlab{}.
\newblock \showarticletitle{Holes in the Outline: Subject-Dependent Abstract
  Quality and Its Implications for Scientific Literature Search}
  \emph{(\bibinfo{series}{CHIIR '19})}. \bibinfo{publisher}{Association for
  Computing Machinery}, \bibinfo{address}{New York, NY, USA},
  \bibinfo{pages}{289–293}.
\newblock
\showISBNx{9781450360258}
\urldef\tempurl%
\url{https://doi.org/10.1145/3295750.3298953}
\showDOI{\tempurl}


\bibitem[\protect\citeauthoryear{Jacovi, Niu, Goldberg, and Sugiyama}{Jacovi
  et~al\mbox{.}}{2021}]%
        {jacovi2021scalable}
\bibfield{author}{\bibinfo{person}{Alon Jacovi}, \bibinfo{person}{Gang Niu},
  \bibinfo{person}{Yoav Goldberg}, {and} \bibinfo{person}{Masashi Sugiyama}.}
  \bibinfo{year}{2021}\natexlab{}.
\newblock \showarticletitle{Scalable Evaluation and Improvement of Document Set
  Expansion via Neural Positive-Unlabeled Learning}. In
  \bibinfo{booktitle}{\emph{Proceedings of the 16th Conference of the European
  Chapter of the Association for Computational Linguistics: Main Volume}}.
  \bibinfo{publisher}{Association for Computational Linguistics},
  \bibinfo{address}{Online}, \bibinfo{pages}{581--592}.
\newblock
\urldef\tempurl%
\url{https://www.aclweb.org/anthology/2021.eacl-main.47}
\showURL{%
\tempurl}


\bibitem[\protect\citeauthoryear{Jain, Banner, van~de Meent, Marshall, and
  Wallace}{Jain et~al\mbox{.}}{2018}]%
        {jain2018disentangled}
\bibfield{author}{\bibinfo{person}{Sarthak Jain}, \bibinfo{person}{Edward
  Banner}, \bibinfo{person}{Jan-Willem van~de Meent}, \bibinfo{person}{Iain~J.
  Marshall}, {and} \bibinfo{person}{Byron~C. Wallace}.}
  \bibinfo{year}{2018}\natexlab{}.
\newblock \showarticletitle{Learning Disentangled Representations of Texts with
  Application to Biomedical Abstracts}. In
  \bibinfo{booktitle}{\emph{Proceedings of the 2018 Conference on Empirical
  Methods in Natural Language Processing}}. \bibinfo{publisher}{Association for
  Computational Linguistics}, \bibinfo{address}{Brussels, Belgium},
  \bibinfo{pages}{4683--4693}.
\newblock
\urldef\tempurl%
\url{https://doi.org/10.18653/v1/D18-1497}
\showDOI{\tempurl}


\bibitem[\protect\citeauthoryear{Jurgens, Kumar, Hoover, McFarland, and
  Jurafsky}{Jurgens et~al\mbox{.}}{2018}]%
        {jurgens2018measuring}
\bibfield{author}{\bibinfo{person}{David Jurgens}, \bibinfo{person}{Srijan
  Kumar}, \bibinfo{person}{Raine Hoover}, \bibinfo{person}{Dan McFarland},
  {and} \bibinfo{person}{Dan Jurafsky}.} \bibinfo{year}{2018}\natexlab{}.
\newblock \showarticletitle{Measuring the Evolution of a Scientific Field
  through Citation Frames}.
\newblock \bibinfo{journal}{\emph{Transactions of the Association for
  Computational Linguistics}}  \bibinfo{volume}{6} (\bibinfo{year}{2018}),
  \bibinfo{pages}{391--406}.
\newblock
\urldef\tempurl%
\url{https://doi.org/10.1162/tacl_a_00028}
\showDOI{\tempurl}


\bibitem[\protect\citeauthoryear{Kamper, Anastassiou, and Livescu}{Kamper
  et~al\mbox{.}}{2019}]%
        {kamper2019musicqbe}
\bibfield{author}{\bibinfo{person}{Herman Kamper}, \bibinfo{person}{Aristotelis
  Anastassiou}, {and} \bibinfo{person}{Karen Livescu}.}
  \bibinfo{year}{2019}\natexlab{}.
\newblock \showarticletitle{Semantic Query-by-example Speech Search Using
  Visual Grounding}. In \bibinfo{booktitle}{\emph{ICASSP 2019 - 2019 IEEE
  International Conference on Acoustics, Speech and Signal Processing
  (ICASSP)}}. \bibinfo{pages}{7120--7124}.
\newblock
\urldef\tempurl%
\url{https://doi.org/10.1109/ICASSP.2019.8683275}
\showDOI{\tempurl}


\bibitem[\protect\citeauthoryear{Karimzadehgan, Zhai, and
  Belford}{Karimzadehgan et~al\mbox{.}}{2008}]%
        {Karimzadehgan2008aspectreview}
\bibfield{author}{\bibinfo{person}{Maryam Karimzadehgan},
  \bibinfo{person}{ChengXiang Zhai}, {and} \bibinfo{person}{Geneva Belford}.}
  \bibinfo{year}{2008}\natexlab{}.
\newblock \showarticletitle{Multi-Aspect Expertise Matching for Review
  Assignment}. In \bibinfo{booktitle}{\emph{Proceedings of the 17th ACM
  Conference on Information and Knowledge Management}}
  \emph{(\bibinfo{series}{CIKM '08})}. \bibinfo{publisher}{Association for
  Computing Machinery}, \bibinfo{address}{New York, NY, USA},
  \bibinfo{pages}{1113–1122}.
\newblock
\showISBNx{9781595939913}
\urldef\tempurl%
\url{https://doi.org/10.1145/1458082.1458230}
\showDOI{\tempurl}


\bibitem[\protect\citeauthoryear{Kobayashi, Shimbo, and Matsumoto}{Kobayashi
  et~al\mbox{.}}{2018}]%
        {kobayashi2018citation}
\bibfield{author}{\bibinfo{person}{Yuta Kobayashi}, \bibinfo{person}{Masashi
  Shimbo}, {and} \bibinfo{person}{Yuji Matsumoto}.}
  \bibinfo{year}{2018}\natexlab{}.
\newblock \showarticletitle{Citation Recommendation Using Distributed
  Representation of Discourse Facets in Scientific Articles}. In
  \bibinfo{booktitle}{\emph{Proceedings of the 18th ACM/IEEE on Joint
  Conference on Digital Libraries}} \emph{(\bibinfo{series}{JCDL '18})}.
  \bibinfo{publisher}{Association for Computing Machinery},
  \bibinfo{address}{New York, NY, USA}, \bibinfo{pages}{243–251}.
\newblock
\showISBNx{9781450351782}
\urldef\tempurl%
\url{https://doi.org/10.1145/3197026.3197059}
\showDOI{\tempurl}


\bibitem[\protect\citeauthoryear{Kong}{Kong}{2016a}]%
        {weize2016facetedsearchthesis}
\bibfield{author}{\bibinfo{person}{Weize Kong}.}
  \bibinfo{year}{2016}\natexlab{a}.
\newblock \showarticletitle{Extending Faceted Search to the Open-Domain Web}.
\newblock \bibinfo{journal}{\emph{SIGIR Forum}} \bibinfo{volume}{50},
  \bibinfo{number}{1} (\bibinfo{date}{June} \bibinfo{year}{2016}),
  \bibinfo{pages}{90–91}.
\newblock
\showISSN{0163-5840}
\urldef\tempurl%
\url{https://doi.org/10.1145/2964797.2964814}
\showDOI{\tempurl}


\bibitem[\protect\citeauthoryear{Kong}{Kong}{2016b}]%
        {Kong2016thesis}
\bibfield{author}{\bibinfo{person}{Weize Kong}.}
  \bibinfo{year}{2016}\natexlab{b}.
\newblock \showarticletitle{Extending Faceted Search to the Open-Domain Web}.
\newblock \bibinfo{journal}{\emph{SIGIR Forum}} \bibinfo{volume}{50},
  \bibinfo{number}{1} (\bibinfo{date}{June} \bibinfo{year}{2016}),
  \bibinfo{pages}{90–91}.
\newblock
\showISSN{0163-5840}
\urldef\tempurl%
\url{https://doi.org/10.1145/2964797.2964814}
\showDOI{\tempurl}


\bibitem[\protect\citeauthoryear{Kong and Allan}{Kong and Allan}{2016}]%
        {Kong2016precision}
\bibfield{author}{\bibinfo{person}{Weize Kong} {and} \bibinfo{person}{James
  Allan}.} \bibinfo{year}{2016}\natexlab{}.
\newblock \showarticletitle{Precision-Oriented Query Facet Extraction}. In
  \bibinfo{booktitle}{\emph{Proceedings of the 25th ACM International on
  Conference on Information and Knowledge Management}}
  \emph{(\bibinfo{series}{CIKM '16})}. \bibinfo{publisher}{Association for
  Computing Machinery}, \bibinfo{address}{New York, NY, USA},
  \bibinfo{pages}{1433–1442}.
\newblock
\showISBNx{9781450340731}
\urldef\tempurl%
\url{https://doi.org/10.1145/2983323.2983824}
\showDOI{\tempurl}


\bibitem[\protect\citeauthoryear{Kruiper, Chen-Burger, and Desmulliez}{Kruiper
  et~al\mbox{.}}{2016}]%
        {Kruiper2016biomimetics}
\bibfield{author}{\bibinfo{person}{Ruben Kruiper}, \bibinfo{person}{Jessica
  Chen-Burger}, {and} \bibinfo{person}{Marc P.~Y. Desmulliez}.}
  \bibinfo{year}{2016}\natexlab{}.
\newblock \showarticletitle{Computer-Aided Biomimetics}. In
  \bibinfo{booktitle}{\emph{Biomimetic and Biohybrid Systems}},
  \bibfield{editor}{\bibinfo{person}{Nathan~F. Lepora}, \bibinfo{person}{Anna
  Mura}, \bibinfo{person}{Michael Mangan}, \bibinfo{person}{Paul~F.M.J.
  Verschure}, \bibinfo{person}{Marc Desmulliez}, {and} \bibinfo{person}{Tony~J.
  Prescott}} (Eds.). \bibinfo{publisher}{Springer International Publishing},
  \bibinfo{address}{Cham}, \bibinfo{pages}{131--143}.
\newblock
\showISBNx{978-3-319-42417-0}
\urldef\tempurl%
\url{https://doi.org/10.1007/978-3-319-42417-0_13}
\showDOI{\tempurl}


\bibitem[\protect\citeauthoryear{Ksikes}{Ksikes}{2014}]%
        {ksikes2014towards}
\bibfield{author}{\bibinfo{person}{Alex Ksikes}.}
  \bibinfo{year}{2014}\natexlab{}.
\newblock \emph{\bibinfo{title}{Towards exploratory faceted search systems}}.
\newblock \bibinfo{thesistype}{Ph.D. Dissertation}. \bibinfo{school}{University
  of Cambridge}.
\newblock
\urldef\tempurl%
\url{https://doi.org/10.17863/CAM.14080}
\showDOI{\tempurl}


\bibitem[\protect\citeauthoryear{Lafia}{Lafia}{2020}]%
        {lafia2020designing}
\bibfield{author}{\bibinfo{person}{Sara~Katherine Lafia}.}
  \bibinfo{year}{2020}\natexlab{}.
\newblock \emph{\bibinfo{title}{Designing for Serendipity: Research Data
  Curation in Topic Spaces}}.
\newblock \bibinfo{thesistype}{Ph.D. Dissertation}. \bibinfo{school}{UC Santa
  Barbara}.
\newblock
\urldef\tempurl%
\url{https://escholarship.org/uc/item/5647q82f}
\showURL{%
\tempurl}


\bibitem[\protect\citeauthoryear{Landhuis}{Landhuis}{2016}]%
        {landhuis2016infoverload}
\bibfield{author}{\bibinfo{person}{Esther Landhuis}.}
  \bibinfo{year}{2016}\natexlab{}.
\newblock \showarticletitle{Scientific literature: Information overload}.
\newblock \bibinfo{journal}{\emph{Nature}} \bibinfo{volume}{535},
  \bibinfo{number}{7612} (\bibinfo{year}{2016}), \bibinfo{pages}{457--458}.
\newblock
\urldef\tempurl%
\url{https://doi.org/10.1038/nj7612-457a}
\showDOI{\tempurl}


\bibitem[\protect\citeauthoryear{Lavrac, Martinc, Pollak, and Cestnik}{Lavrac
  et~al\mbox{.}}{2020}]%
        {Lavrac2020BisociativeLD}
\bibfield{author}{\bibinfo{person}{N. Lavrac}, \bibinfo{person}{Matej Martinc},
  \bibinfo{person}{Senja Pollak}, {and} \bibinfo{person}{B. Cestnik}.}
  \bibinfo{year}{2020}\natexlab{}.
\newblock \showarticletitle{Bisociative Literature-Based Discovery: Lessons
  Learned and New Prospects}. In \bibinfo{booktitle}{\emph{ICCC}}.
\newblock
\urldef\tempurl%
\url{http://computationalcreativity.net/iccc20/papers/034-iccc20.pdf}
\showURL{%
\tempurl}


\bibitem[\protect\citeauthoryear{Lissandrini, Mottin, Palpanas, and
  Velegrakis}{Lissandrini et~al\mbox{.}}{2019}]%
        {lissandrini2019exampletutorial}
\bibfield{author}{\bibinfo{person}{Matteo Lissandrini}, \bibinfo{person}{Davide
  Mottin}, \bibinfo{person}{Themis Palpanas}, {and} \bibinfo{person}{Yannis
  Velegrakis}.} \bibinfo{year}{2019}\natexlab{}.
\newblock \showarticletitle{Example-Based Search: A New Frontier for
  Exploratory Search}. In \bibinfo{booktitle}{\emph{Proceedings of the 42nd
  International ACM SIGIR Conference on Research and Development in Information
  Retrieval}} \emph{(\bibinfo{series}{SIGIR'19})}.
  \bibinfo{publisher}{Association for Computing Machinery},
  \bibinfo{address}{New York, NY, USA}, \bibinfo{pages}{1411–1412}.
\newblock
\showISBNx{9781450361729}
\urldef\tempurl%
\url{https://doi.org/10.1145/3331184.3331387}
\showDOI{\tempurl}


\bibitem[\protect\citeauthoryear{Lo, Wang, Neumann, Kinney, and Weld}{Lo
  et~al\mbox{.}}{2020}]%
        {lo2020s2orc}
\bibfield{author}{\bibinfo{person}{Kyle Lo}, \bibinfo{person}{Lucy~Lu Wang},
  \bibinfo{person}{Mark Neumann}, \bibinfo{person}{Rodney Kinney}, {and}
  \bibinfo{person}{Daniel Weld}.} \bibinfo{year}{2020}\natexlab{}.
\newblock \showarticletitle{{S}2{ORC}: The Semantic Scholar Open Research
  Corpus}. In \bibinfo{booktitle}{\emph{Proceedings of the 58th Annual Meeting
  of the Association for Computational Linguistics}}.
  \bibinfo{publisher}{Association for Computational Linguistics},
  \bibinfo{address}{Online}, \bibinfo{pages}{4969--4983}.
\newblock
\urldef\tempurl%
\url{https://doi.org/10.18653/v1/2020.acl-main.447}
\showDOI{\tempurl}


\bibitem[\protect\citeauthoryear{Medin, Goldstone, and Gentner}{Medin
  et~al\mbox{.}}{1990}]%
        {douglas1990similarity}
\bibfield{author}{\bibinfo{person}{Douglas~L. Medin},
  \bibinfo{person}{Robert~L. Goldstone}, {and} \bibinfo{person}{Dedre
  Gentner}.} \bibinfo{year}{1990}\natexlab{}.
\newblock \showarticletitle{Similarity Involving Attributes and Relations:
  Judgments of Similarity and Difference Are Not Inverses}.
\newblock \bibinfo{journal}{\emph{Psychological Science}} \bibinfo{volume}{1},
  \bibinfo{number}{1} (\bibinfo{year}{1990}), \bibinfo{pages}{64--69}.
\newblock
\showISSN{09567976, 14679280}
\urldef\tempurl%
\url{http://www.jstor.org/stable/40062393}
\showURL{%
\tempurl}


\bibitem[\protect\citeauthoryear{Mozer, Miratrix, Kaufman, and
  Jason~Anastasopoulos}{Mozer et~al\mbox{.}}{2020}]%
        {mozer2020tm}
\bibfield{author}{\bibinfo{person}{Reagan Mozer}, \bibinfo{person}{Luke
  Miratrix}, \bibinfo{person}{Aaron~Russell Kaufman}, {and} \bibinfo{person}{L.
  Jason~Anastasopoulos}.} \bibinfo{year}{2020}\natexlab{}.
\newblock \showarticletitle{Matching with Text Data: An Experimental Evaluation
  of Methods for Matching Documents and of Measuring Match Quality}.
\newblock \bibinfo{journal}{\emph{Political Analysis}} \bibinfo{volume}{28},
  \bibinfo{number}{4} (\bibinfo{year}{2020}), \bibinfo{pages}{445–468}.
\newblock
\urldef\tempurl%
\url{https://doi.org/10.1017/pan.2020.1}
\showDOI{\tempurl}


\bibitem[\protect\citeauthoryear{Nallapati, Peerreddy, and Singhal}{Nallapati
  et~al\mbox{.}}{2012}]%
        {nallapati2012skierarchy}
\bibfield{author}{\bibinfo{person}{Ramesh Nallapati}, \bibinfo{person}{Sanga
  Peerreddy}, {and} \bibinfo{person}{Prateek Singhal}.}
  \bibinfo{year}{2012}\natexlab{}.
\newblock \bibinfo{booktitle}{\emph{Skierarchy: Extending the power of
  crowdsourcing using a hierarchy of domain experts, crowd and machine
  learning}}.
\newblock \bibinfo{type}{{T}echnical {R}eport}.
\newblock
\urldef\tempurl%
\url{https://apps.dtic.mil/sti/citations/ADA581773}
\showURL{%
\tempurl}


\bibitem[\protect\citeauthoryear{Nanba and Okumura}{Nanba and Okumura}{1999}]%
        {nanba1999towards}
\bibfield{author}{\bibinfo{person}{Hidetsugu Nanba} {and}
  \bibinfo{person}{Manabu Okumura}.} \bibinfo{year}{1999}\natexlab{}.
\newblock \showarticletitle{Towards Multi-Paper Summarization Reference
  Information}. In \bibinfo{booktitle}{\emph{Proceedings of the 16th
  International Joint Conference on Artificial Intelligence - Volume 2}}
  \emph{(\bibinfo{series}{IJCAI'99})}. \bibinfo{publisher}{Morgan Kaufmann
  Publishers Inc.}, \bibinfo{address}{San Francisco, CA, USA},
  \bibinfo{pages}{926–931}.
\newblock


\bibitem[\protect\citeauthoryear{Neves, Butzke, and Grune}{Neves
  et~al\mbox{.}}{2019}]%
        {neves2019evaluation}
\bibfield{author}{\bibinfo{person}{Mariana Neves}, \bibinfo{person}{Daniel
  Butzke}, {and} \bibinfo{person}{Barbara Grune}.}
  \bibinfo{year}{2019}\natexlab{}.
\newblock \showarticletitle{Evaluation of Scientific Elements for Text
  Similarity in Biomedical Publications}. In
  \bibinfo{booktitle}{\emph{Proceedings of the 6th Workshop on Argument
  Mining}}. \bibinfo{publisher}{Association for Computational Linguistics},
  \bibinfo{address}{Florence, Italy}.
\newblock
\urldef\tempurl%
\url{https://www.aclweb.org/anthology/W19-4515}
\showURL{%
\tempurl}


\bibitem[\protect\citeauthoryear{OpenReview}{OpenReview}{[n.d.]}]%
        {ORExpertise}
\bibfield{author}{\bibinfo{person}{OpenReview}.}
  \bibinfo{year}{[n.d.]}\natexlab{}.
\newblock \bibinfo{title}{Paper-reviewer affinity modeling for OpenReview}.
\newblock
  \bibinfo{howpublished}{\url{https://github.com/openreview/openreview-expertise}}.
\newblock
\newblock
\shownote{Accessed: 27 August 2021.}


\bibitem[\protect\citeauthoryear{Ostendorff}{Ostendorff}{2020}]%
        {ostendorff2020contextual}
\bibfield{author}{\bibinfo{person}{Malte Ostendorff}.}
  \bibinfo{year}{2020}\natexlab{}.
\newblock \showarticletitle{Contextual Document Similarity for Content-based
  Literature Recommender Systems}.
\newblock \bibinfo{journal}{\emph{Proceedings of the Doctoral Consortium at
  ACM/IEEE Joint Conference on Digital Libraries}}.
\newblock


\bibitem[\protect\citeauthoryear{Ostendorff, Ruas, Blume, Gipp, and
  Rehm}{Ostendorff et~al\mbox{.}}{2020}]%
        {ostendorff2020aspectcoling}
\bibfield{author}{\bibinfo{person}{Malte Ostendorff}, \bibinfo{person}{Terry
  Ruas}, \bibinfo{person}{Till Blume}, \bibinfo{person}{Bela Gipp}, {and}
  \bibinfo{person}{Georg Rehm}.} \bibinfo{year}{2020}\natexlab{}.
\newblock \showarticletitle{Aspect-based Document Similarity for Research
  Papers}. In \bibinfo{booktitle}{\emph{Proceedings of the 28th International
  Conference on Computational Linguistics}}. \bibinfo{publisher}{International
  Committee on Computational Linguistics}, \bibinfo{address}{Barcelona, Spain
  (Online)}, \bibinfo{pages}{6194--6206}.
\newblock
\urldef\tempurl%
\url{https://doi.org/10.18653/v1/2020.coling-main.545}
\showDOI{\tempurl}


\bibitem[\protect\citeauthoryear{Pain}{Pain}{2016}]%
        {pain2016keep}
\bibfield{author}{\bibinfo{person}{Elisabeth Pain}.}
  \bibinfo{year}{2016}\natexlab{}.
\newblock \showarticletitle{How to keep up with the scientific literature}.
\newblock \bibinfo{journal}{\emph{Science Careers}}  \bibinfo{volume}{30}
  (\bibinfo{year}{2016}).
\newblock
\urldef\tempurl%
\url{https://www.science.org/content/article/how-keep-scientific-literature-rev2}
\showURL{%
\tempurl}


\bibitem[\protect\citeauthoryear{Reimers and Gurevych}{Reimers and
  Gurevych}{2019}]%
        {reimers2019sentencebert}
\bibfield{author}{\bibinfo{person}{Nils Reimers} {and} \bibinfo{person}{Iryna
  Gurevych}.} \bibinfo{year}{2019}\natexlab{}.
\newblock \showarticletitle{Sentence-BERT: Sentence Embeddings using Siamese
  BERT-Networks}. In \bibinfo{booktitle}{\emph{Proceedings of the 2019
  Conference on Empirical Methods in Natural Language Processing}}.
  \bibinfo{publisher}{Association for Computational Linguistics}.
\newblock
\urldef\tempurl%
\url{https://arxiv.org/abs/1908.10084}
\showURL{%
\tempurl}


\bibitem[\protect\citeauthoryear{Roberts, Stewart, and Nielsen}{Roberts
  et~al\mbox{.}}{2020}]%
        {Roberts2020gendertextmat}
\bibfield{author}{\bibinfo{person}{Margaret~E. Roberts},
  \bibinfo{person}{Brandon~M. Stewart}, {and} \bibinfo{person}{Richard~A.
  Nielsen}.} \bibinfo{year}{2020}\natexlab{}.
\newblock \showarticletitle{Adjusting for Confounding with Text Matching}.
\newblock \bibinfo{journal}{\emph{American Journal of Political Science}}
  \bibinfo{volume}{64}, \bibinfo{number}{4} (\bibinfo{year}{2020}),
  \bibinfo{pages}{887--903}.
\newblock
\urldef\tempurl%
\url{https://doi.org/10.1111/ajps.12526}
\showDOI{\tempurl}


\bibitem[\protect\citeauthoryear{Sarwar and Allan}{Sarwar and Allan}{2020a}]%
        {Sarwar2020QBEEvent}
\bibfield{author}{\bibinfo{person}{Sheikh~Muhammad Sarwar} {and}
  \bibinfo{person}{James Allan}.} \bibinfo{year}{2020}\natexlab{a}.
\newblock \showarticletitle{Query by Example for Cross-Lingual Event
  Retrieval}. In \bibinfo{booktitle}{\emph{Proceedings of the 43rd
  International ACM SIGIR Conference on Research and Development in Information
  Retrieval}} \emph{(\bibinfo{series}{SIGIR '20})}.
  \bibinfo{publisher}{Association for Computing Machinery},
  \bibinfo{address}{New York, NY, USA}, \bibinfo{pages}{1601–1604}.
\newblock
\showISBNx{9781450380164}
\urldef\tempurl%
\url{https://doi.org/10.1145/3397271.3401283}
\showDOI{\tempurl}


\bibitem[\protect\citeauthoryear{Sarwar and Allan}{Sarwar and Allan}{2020b}]%
        {sarwar2020queryevent}
\bibfield{author}{\bibinfo{person}{Sheikh~Muhammad Sarwar} {and}
  \bibinfo{person}{James Allan}.} \bibinfo{year}{2020}\natexlab{b}.
\newblock \showarticletitle{Query by Example for Cross-Lingual Event
  Retrieval}. In \bibinfo{booktitle}{\emph{Proceedings of the 43rd
  International ACM SIGIR Conference on Research and Development in Information
  Retrieval}} \emph{(\bibinfo{series}{SIGIR '20})}.
  \bibinfo{publisher}{Association for Computing Machinery},
  \bibinfo{address}{New York, NY, USA}, \bibinfo{pages}{1601–1604}.
\newblock
\showISBNx{9781450380164}
\urldef\tempurl%
\url{https://doi.org/10.1145/3397271.3401283}
\showDOI{\tempurl}


\bibitem[\protect\citeauthoryear{Scholar}{Scholar}{[n.d.]}]%
        {S2Recommendation}
\bibfield{author}{\bibinfo{person}{Semantic Scholar}.}
  \bibinfo{year}{[n.d.]}\natexlab{}.
\newblock \bibinfo{title}{Semantic Scholar on Twitter}.
\newblock
  \bibinfo{howpublished}{\url{https://twitter.com/SemanticScholar/status/1267867735318355968/retweets}}.
\newblock
\newblock
\shownote{Accessed: 27 August 2021.}


\bibitem[\protect\citeauthoryear{Shlain, Taub-Tabib, Sadde, and
  Goldberg}{Shlain et~al\mbox{.}}{2020}]%
        {Shlain2020spike}
\bibfield{author}{\bibinfo{person}{Micah Shlain}, \bibinfo{person}{Hillel
  Taub-Tabib}, \bibinfo{person}{Shoval Sadde}, {and} \bibinfo{person}{Yoav
  Goldberg}.} \bibinfo{year}{2020}\natexlab{}.
\newblock \showarticletitle{Syntactic Search by Example}. In
  \bibinfo{booktitle}{\emph{Proceedings of the 58th Annual Meeting of the
  Association for Computational Linguistics: System Demonstrations}}.
  \bibinfo{publisher}{Association for Computational Linguistics},
  \bibinfo{address}{Online}, \bibinfo{pages}{17--23}.
\newblock
\urldef\tempurl%
\url{https://doi.org/10.18653/v1/2020.acl-demos.3}
\showDOI{\tempurl}


\bibitem[\protect\citeauthoryear{Taub~Tabib, Shlain, Sadde, Lahav, Eyal, Cohen,
  and Goldberg}{Taub~Tabib et~al\mbox{.}}{2020}]%
        {tabib2020interactive}
\bibfield{author}{\bibinfo{person}{Hillel Taub~Tabib}, \bibinfo{person}{Micah
  Shlain}, \bibinfo{person}{Shoval Sadde}, \bibinfo{person}{Dan Lahav},
  \bibinfo{person}{Matan Eyal}, \bibinfo{person}{Yaara Cohen}, {and}
  \bibinfo{person}{Yoav Goldberg}.} \bibinfo{year}{2020}\natexlab{}.
\newblock \showarticletitle{Interactive Extractive Search over Biomedical
  Corpora}. In \bibinfo{booktitle}{\emph{Proceedings of the 19th SIGBioMed
  Workshop on Biomedical Language Processing}}. \bibinfo{publisher}{Association
  for Computational Linguistics}, \bibinfo{address}{Online},
  \bibinfo{pages}{28--37}.
\newblock
\urldef\tempurl%
\url{https://doi.org/10.18653/v1/2020.bionlp-1.3}
\showDOI{\tempurl}


\bibitem[\protect\citeauthoryear{Teufel, Siddharthan, and Tidhar}{Teufel
  et~al\mbox{.}}{2006}]%
        {teufel2006automatic}
\bibfield{author}{\bibinfo{person}{Simone Teufel}, \bibinfo{person}{Advaith
  Siddharthan}, {and} \bibinfo{person}{Dan Tidhar}.}
  \bibinfo{year}{2006}\natexlab{}.
\newblock \showarticletitle{Automatic classification of citation function}. In
  \bibinfo{booktitle}{\emph{Proceedings of the 2006 Conference on Empirical
  Methods in Natural Language Processing}}. \bibinfo{publisher}{Association for
  Computational Linguistics}, \bibinfo{address}{Sydney, Australia},
  \bibinfo{pages}{103--110}.
\newblock
\urldef\tempurl%
\url{https://aclanthology.org/W06-1613}
\showURL{%
\tempurl}


\bibitem[\protect\citeauthoryear{Tunkelang}{Tunkelang}{2006}]%
        {tunkelang2006dynamic}
\bibfield{author}{\bibinfo{person}{Daniel Tunkelang}.}
  \bibinfo{year}{2006}\natexlab{}.
\newblock \showarticletitle{Dynamic category sets: An approach for faceted
  search}. In \bibinfo{booktitle}{\emph{ACM SIGIR}}, Vol.~\bibinfo{volume}{6}.
\newblock
\urldef\tempurl%
\url{https://citeseerx.ist.psu.edu/viewdoc/download?doi=10.1.1.83.4202&rep=rep1&type=pdf}
\showURL{%
\tempurl}


\bibitem[\protect\citeauthoryear{Upadhyay, Bedathur, Chakraborty, and
  Ramanath}{Upadhyay et~al\mbox{.}}{2020}]%
        {Upadhyay2020AspectKB}
\bibfield{author}{\bibinfo{person}{Prajna Upadhyay}, \bibinfo{person}{Srikanta
  Bedathur}, \bibinfo{person}{Tanmoy Chakraborty}, {and} \bibinfo{person}{Maya
  Ramanath}.} \bibinfo{year}{2020}\natexlab{}.
\newblock \showarticletitle{Aspect-Based Academic Search Using Domain-Specific
  KB}. In \bibinfo{booktitle}{\emph{Advances in Information Retrieval}},
  \bibfield{editor}{\bibinfo{person}{Joemon~M. Jose}, \bibinfo{person}{Emine
  Yilmaz}, \bibinfo{person}{Jo{\~a}o Magalh{\~a}es}, \bibinfo{person}{Pablo
  Castells}, \bibinfo{person}{Nicola Ferro}, \bibinfo{person}{M{\'a}rio~J.
  Silva}, {and} \bibinfo{person}{Fl{\'a}vio Martins}} (Eds.).
  \bibinfo{publisher}{Springer International Publishing},
  \bibinfo{address}{Cham}, \bibinfo{pages}{418--424}.
\newblock
\showISBNx{978-3-030-45442-5}
\urldef\tempurl%
\url{https://link.springer.com/chapter/10.1007/978-3-030-45442-5_52}
\showURL{%
\tempurl}


\bibitem[\protect\citeauthoryear{Vanschoren, van Rijn, Bischl, and
  Torgo}{Vanschoren et~al\mbox{.}}{2014}]%
        {Vanschoren2013OpenML}
\bibfield{author}{\bibinfo{person}{Joaquin Vanschoren}, \bibinfo{person}{Jan~N.
  van Rijn}, \bibinfo{person}{Bernd Bischl}, {and} \bibinfo{person}{Luis
  Torgo}.} \bibinfo{year}{2014}\natexlab{}.
\newblock \showarticletitle{OpenML: Networked Science in Machine Learning}.
\newblock \bibinfo{journal}{\emph{SIGKDD Explor. Newsl.}} \bibinfo{volume}{15},
  \bibinfo{number}{2} (\bibinfo{date}{June} \bibinfo{year}{2014}),
  \bibinfo{pages}{49–60}.
\newblock
\showISSN{1931-0145}
\urldef\tempurl%
\url{https://doi.org/10.1145/2641190.2641198}
\showURL{%
\tempurl}


\bibitem[\protect\citeauthoryear{Voorhees, Alam, Bedrick, Demner-Fushman,
  Hersh, Lo, Roberts, Soboroff, and Wang}{Voorhees et~al\mbox{.}}{2021}]%
        {treccovid2021}
\bibfield{author}{\bibinfo{person}{Ellen Voorhees}, \bibinfo{person}{Tasmeer
  Alam}, \bibinfo{person}{Steven Bedrick}, \bibinfo{person}{Dina
  Demner-Fushman}, \bibinfo{person}{William~R. Hersh}, \bibinfo{person}{Kyle
  Lo}, \bibinfo{person}{Kirk Roberts}, \bibinfo{person}{Ian Soboroff}, {and}
  \bibinfo{person}{Lucy~Lu Wang}.} \bibinfo{year}{2021}\natexlab{}.
\newblock \showarticletitle{TREC-COVID: Constructing a Pandemic Information
  Retrieval Test Collection}.
\newblock \bibinfo{journal}{\emph{SIGIR Forum}}, Article \bibinfo{articleno}{1}
  (\bibinfo{date}{Feb.} \bibinfo{year}{2021}), \bibinfo{numpages}{12}~pages.
\newblock
\urldef\tempurl%
\url{https://doi.org/10.1145/3451964.3451965}
\showURL{%
\tempurl}


\bibitem[\protect\citeauthoryear{Wadden, Lin, Lo, Wang, van Zuylen, Cohan, and
  Hajishirzi}{Wadden et~al\mbox{.}}{2020}]%
        {wadden2020fact}
\bibfield{author}{\bibinfo{person}{David Wadden}, \bibinfo{person}{Shanchuan
  Lin}, \bibinfo{person}{Kyle Lo}, \bibinfo{person}{Lucy~Lu Wang},
  \bibinfo{person}{Madeleine van Zuylen}, \bibinfo{person}{Arman Cohan}, {and}
  \bibinfo{person}{Hannaneh Hajishirzi}.} \bibinfo{year}{2020}\natexlab{}.
\newblock \showarticletitle{Fact or Fiction: Verifying Scientific Claims}. In
  \bibinfo{booktitle}{\emph{Proceedings of the 2020 Conference on Empirical
  Methods in Natural Language Processing (EMNLP)}}.
  \bibinfo{publisher}{Association for Computational Linguistics},
  \bibinfo{address}{Online}, \bibinfo{pages}{7534--7550}.
\newblock
\urldef\tempurl%
\url{https://doi.org/10.18653/v1/2020.emnlp-main.609}
\showDOI{\tempurl}


\bibitem[\protect\citeauthoryear{Wang, Wang, Li, He, and Liu}{Wang
  et~al\mbox{.}}{2013}]%
        {wang13ndcg}
\bibfield{author}{\bibinfo{person}{Yining Wang}, \bibinfo{person}{Liwei Wang},
  \bibinfo{person}{Yuanzhi Li}, \bibinfo{person}{Di He}, {and}
  \bibinfo{person}{Tie-Yan Liu}.} \bibinfo{year}{2013}\natexlab{}.
\newblock \showarticletitle{A Theoretical Analysis of NDCG Type Ranking
  Measures}. In \bibinfo{booktitle}{\emph{Proceedings of the 26th Annual
  Conference on Learning Theory}} \emph{(\bibinfo{series}{Proceedings of
  Machine Learning Research})}, \bibfield{editor}{\bibinfo{person}{Shai
  Shalev-Shwartz} {and} \bibinfo{person}{Ingo Steinwart}} (Eds.),
  Vol.~\bibinfo{volume}{30}. \bibinfo{publisher}{PMLR},
  \bibinfo{address}{Princeton, NJ, USA}, \bibinfo{pages}{25--54}.
\newblock
\urldef\tempurl%
\url{https://proceedings.mlr.press/v30/Wang13.html}
\showURL{%
\tempurl}


\bibitem[\protect\citeauthoryear{Zhu and Wu}{Zhu and Wu}{2014}]%
        {zhu2014multiqbe}
\bibfield{author}{\bibinfo{person}{Mingzhu Zhu} {and}
  \bibinfo{person}{Yi-Fang~Brook Wu}.} \bibinfo{year}{2014}\natexlab{}.
\newblock \showarticletitle{Search by Multiple Examples}. In
  \bibinfo{booktitle}{\emph{Proceedings of the 7th ACM International Conference
  on Web Search and Data Mining}} \emph{(\bibinfo{series}{WSDM '14})}.
  \bibinfo{publisher}{Association for Computing Machinery},
  \bibinfo{address}{New York, NY, USA}, \bibinfo{pages}{667–672}.
\newblock
\showISBNx{9781450323512}
\urldef\tempurl%
\url{https://doi.org/10.1145/2556195.2556206}
\showDOI{\tempurl}


\end{thebibliography}

\newpage
\appendix
\section{Checklist Materials}
\begin{description}
    \item Datasheet: \url{https://github.com/iesl/CSFCube/blob/master/datasheet.md}
    \item License: \url{https://github.com/iesl/CSFCube/blob/master/LICENSE.md}
\end{description}

\section{Full Text vs Abstract Annotations}
\label{appendix-ft-abs}
As we note in \S\ref{sec-rel-analysis}, to examine the effect of annotating the abstract of papers instead of full-text of papers we also conduct a small scale study to examine the scale of differences between the relevance ratings produced each of them. This is done by a single expert annotator annotating relevance based on abstract and full-text separately for 9 query abstracts (3 from each facet) and 5 candidate abstracts each (45 pairs). In making these annotations, queries were picked to ensure all paper-types (\S\ref{sec-datasetdescr}) were represented and candidates were chosen at random from across all relevance levels. Next abstract based relevances were labelled, following this full-text relevances were labelled. In labelling full-text relevances care was taken to not show abstract based relevances or the abstract text. In making full-text judgements the paper was skimmed for content relevant to the target facet rather than read exhaustively. Every full-text judgement pair took about 5 minutes to complete. Finally, while the presented study isnt intended to be statistically robust we believe it presents a reasonable pilot study in support of the abstract based ratings adopted in our dataset annotation.

\section{Baseline Methods}
\label{appendix-baseline-descr}
The methods we choose to evaluate capture a range of granularities and nature of methods: term based methods, pre-trained model based sentence representations, and whole abstract representation models. Note that some of the methods evaluated are included in our set of methods to construct candidate pools, but as noted in \S\ref{sec-datasetdescr-cpooling} they used unfaceted representations.
\begin{description}[noitemsep,nolistsep]
    \item \texttt{fabs\_tfidf}: This is a simple faceted baseline which
builds a sparse TF-IDF representation for the sentences corresponding to the
query facet in the query abstract. Candidates are represented by their whole
abstract TF-IDF representations.
    \item \texttt{fabs\_bm25}: This represents a baseline identical to \texttt{fabs\_tfidf} while using the Okapi \textsc{bm25} weighting scheme.\footnote{\textsc{bm25} implementation: \url{https://github.com/dorianbrown/rank_bm25}}
    \item \texttt{fabs\_cbow200}: This is a dense bag-of-words representation for the sentences corresponding to the
query facet in the query abstract -- token embeddings are averaged. As above, 
candidates are represented by all abstract sentences. The \texttt{word2vec}
embeddings are trained on $800,000$ abstracts from the S2ORC corpus
(\S\ref{sec-datasetdescr}). We used 200 dimensional word embeddings.
    \item \texttt{fabs\_tfidfcbow200}: This baseline combines the above
baselines where the \texttt{word2vec} representations are weighted by TF-IDF
weights prior to being averaged.
    \item \texttt{SentBERT}: SentBERT \citep{reimers2019sentencebert} represents a sentence level model. In our setup we encode all query facet sentences and all candidate abstract sentences individually with SentBERT, and then use the maximum pairwise sentence cosine similarity between the query and candidate sentences to rank candidates. We evaluate two versions of SentBERT, one fine-tuned only on Natural Language Inference (NLI) datasets as in \citet{reimers2019sentencebert} and a second model fine-tuned on NLI and a wide variety of paraphrase text. We term these \texttt{SentBERT-NLI} and \texttt{SentBERT-PP}.\footnote{we use the \texttt{sentence\_transformers} package. In this package, \texttt{SentBERT-NLI} corresponds to \texttt{nli-roberta-base-v2} and \texttt{SentBERT-PP} to \texttt{paraphrase-TinyBERT-L6-v2}.} We choose to use \texttt{SentBERT-PP} given its strong performance on the SciDOCS benchmark \citep{cohan2020specter}.\footnote{Model performances: \url{https://www.sbert.net/docs/pretrained_models.html}}
    \item \texttt{SimCSE}: SimCSE \citep{gao2021simcse}, represents a very recent state of the art sentence similarity model trained in two ways -- an unsupervised manner training an encoder to maximise similarity with a a "dropped-out" representation of the same sentence, and a supervised version trained on NLI data. We denote these as \texttt{UnSimCSE} and \texttt{SuSimCSE}. We use the models \texttt{princeton-nlp/unsup-simcse-bert-base-uncased} and \texttt{princeton-nlp/sup-simcse-bert-base-uncased} made available through the Hugging Face\footnote{\url{https://huggingface.co/}} package.
    \item \textsc{specter}: This approach represents a multi-layer transformer
based SciBERT model fine-tuned on citation network data
\citep{cohan2020specter}. \textsc{specter} operates on titles and the whole abstract of
the papers and represents an entirely un-faceted model. Both queries and
candidates are represented by their \textsc{specter} embeddings. Note that \textsc{specter} was trained on a corpus of randomly selected scientific documents. We also re-implement and train a version of \textsc{specter} on about 660k computer science paper triples with identical hyper-parameters to \textsc{specter}, we call this in-domain model \textsc{specter-id}.
\end{description}
In re-ranking the candidate pool for every query, the L2 distance between the
query and candidate vectors was used unless specified otherwise.

\section{Evaluating Citations}
\label{appendix-cite-eval}
Because we included cited papers in our candidate pools, and manually assign relevance judgments for them, this dataset allows us to examine the common assumption that cited paper abstracts will be relevant to a query paper abstract.
In this analysis, we find cited papers to pre-dominantly be rated at  0 or 1 levels of relevance, 79\% of the times for \texttt{background}, 88\% of the times for
\texttt{method}, and 86\% of the times for \texttt{result}. Given that
citations are often considered incidental signals from which to train models
and often to evaluate them as well, we believe these observations will have
implications for future modeling and evaluation work. We hope future work will
use citations with caution, particularly in evaluation setups for tasks similar to
the one posed here.

\section{Potential Applications}
\label{appendix-applications}
A range of important applications rely on computing similarity between scientific texts. Given that our dataset allow evaluation of document similarity methods in general we believe our test collection fills an important gap in the development and benchmarking of methods intended for these applications.
\begin{description}[nolistsep, noitemsep]
    \item[Exploratory Search:] Content based search with paradigms such as Query by Example has  been considered more suited to exploratory search tasks \citep{ksikes2014towards, Dimitriadou2014EBE, lissandrini2019exampletutorial} than keyword based search. Recent work has also seen AI powered literature navigation tools leveraging content based search at varying levels of granularity \citep{fadaee2020new}. We believe our task formulation directly suits this and will allow development of methods intended for these applications.
    \item[Patent Search:] \citet{Hain2020TextbasedTS}, highlight the case of measuring technological similarities between patents based on the abstracts of patents, and the subsequent employment of this information in mapping patent quality and mapping technological change. Further they also highlight the lack of benchmarks for the measurement of technological similarity between patents \cite[Sec 4.3]{Hain2020TextbasedTS}. While differing in domain we believe our work provides a valuable resource for model development.
    % \item[Precision Medicine:] \citet{Chung2009} highlights the case of  While differing in domain from the presented resource
    \item[Text Matching for Causal Inference:] \citet{mozer2020tm} highlight the importance of text matching for causal inference from observational text data: ``matched  documents  can  be  used  to  make  unbiased  comparisons  between  groups  on external features such as rates of citation''. \citet{Roberts2020gendertextmat}, demonstrate just such a investigation into the gendered biases of citation patterns. The reliance of these analysis on document similarity and matching across a corpus along specific aspects allows our dataset to be of value in developing methods for document matching.
    \item[Expert Search:] Expert search presents an important application, specially in the contex of peer review, where scientific papers must be matched to experts suited to review it. This often involves computing scientific document similarity \citep{berger2020effective}, a venue where our work proves valuable. In the case of work such as  \citet{Karimzadehgan2008aspectreview}, which attempts to find experts along all aspects of a scientific paper, our work provides an even stronger resource.
\end{description}

\section{Extended results}
\label{appendix-extended-results}
\begin{table*}[t]
\centering
\caption{Extended test set results for the set of baselines methods. Metrics (R-Precision, Precision and Recall at 20, NDCG$_{\%20}$, NDCG$_{\%100}$) are computed based on a 2-fold cross-validation, represent averages over per-query metrics, and are reported as percentages. SPECTER-ID performance is reported over three training re-runs, the remaining baselines are reported based on a single set of model parameters released by the respective authors.}
\scalebox{0.7}{
\begin{tabular}{rccccc|ccccc}
                     & \multicolumn{5}{c}{\texttt{background}}   & \multicolumn{5}{c}{\texttt{method}}        \\ 
                     & RP     & P@20   & R@20   & NDCG$_{\%100}$ & NDCG$_{\%20}$   & RP     & P@20   & R@20   & NDCG$_{\%100}$   & NDCG$_{\%20}$ \\\toprule
\texttt{fabs\_tfidf}          & 23.35 & 27.19 & 45.80 & 78.14 & 57.97 & 09.30  & 09.83 & 34.75 & 57.87 & 31.20\\
\texttt{fabs\_bm25}           & 20.12 & 27.81 & 49.85 & 79.02 & 59.39 & 09.37 & 11.63 & 38.29 & 60.68 & 34.59\\
\texttt{fabs\_cbow200}        & 19.61 & 15.94 & 27.97 & 67.68 & 36.56 & 08.65 & 08.33 & 15.69 & 51.58 & 21.14\\
\texttt{fabs\_tfidfcbow200}   & 15.92 & 16.87 & 27.76 & 69.77 & 40.51 & 07.99 & 06.01 & 17.71 & 51.87 & 21.70\\
\texttt{SentBERT-PP}          & 21.24 & 28.75 & 46.67 & 79.14 & 60.80 & 10.00 & 10.83 & 36.30 & 59.50 & 33.40\\
\texttt{SentBERT-NLI}         & 19.02 & 25.00 & 40.13 & 75.80 & 54.23 & 09.11 & 11.46 & 02.89 & 58.52 & 31.10\\
\texttt{UnSimCSE-BERT}        & 18.15 & 23.44 & 36.05 & 74.34 & 51.59 & 08.86 & 09.65 & 27.92 & 59.21 & 31.23\\
\texttt{SuSimCSE-BERT}        & 19.22 & 22.81 & 46.75 & 76.70 & 55.22 & 08.58 & 09.76 & 29.01 & 58.54 & 30.88\\
\textsc{specter}              & \textbf{24.81} & \textbf{35.31} & \textbf{57.45} & {82.24} & 66.70 & {\bf 11.72} & {13.58} & {40.81} & {62.77} & 37.41\\
\textsc{specter-id}              & $\underset{\pm1.3}{24.55}$ & $\underset{\pm0.5}{34.17}$ & $\underset{\pm0.3}{53.26}$ & $\underset{\pm0.8}{\bf 84.31}$ & $\underset{\pm1.71}{\bf 69.22}$ & $\underset{\pm0.3}{10.53}$ & $\underset{\pm1.21}{\bf 16.22}$ & $\underset{\pm3.6}{\bf 44.59}$ & $\underset{\pm0.4}{\bf 64.63}$ & $\underset{\pm0.78}{\bf 42.76}$\\
\multicolumn{1}{l}{} & \multicolumn{5}{c}{\texttt{result}}       & \multicolumn{5}{c}{\texttt{all}}           \\ 
                     & RP     & P@20   & R@20   & NDCG$_{\%100}$ & NDCG$_{\%20}$  & RP     & P@20   & R@20   & NDCG$_{\%100}$   & NDCG$_{\%20}$\\\toprule
\texttt{fabs\_tfidf}   & 11.35 & 16.28 & 38.57 & 66.12 & 41.24 & 14.59 & 17.64 & 39.69 & 67.17 & 43.19\\
\texttt{fabs\_bm25}           & 11.31 & 20.00 & 40.40 & 67.87 & 45.07 & 13.50 & 19.69 & 42.73 & 68.97 & 46.06\\
\texttt{fabs\_cbow200}        & 11.16 & 10.42 & 23.44 & 60.22 & 30.93 & 13.08 & 11.47 & 22.23 & 59.64 & 29.36\\
\texttt{fabs\_tfidfcbow200}   & 10.43 & 10.69 & 24.39 & 60.30 & 32.79 & 11.38 & 11.09 & 23.13 & 60.42 & 31.42\\
\texttt{SentBERT-PP}   & 13.60 & 19.83 & 41.73 & 71.90 & 52.35 & 14.83 & 19.62 & 41.41 & 69.98 & 48.57\\
\texttt{SentBERT-NLI}         & 14.23 & 22.05 & 46.99 & 72.13 & 51.30 & 14.04 & 19.42 & 38.67 & 68.68 & 45.39\\
\texttt{UnSimCSE-BERT}        & 12.00 & 19.58 & 38.95 & 68.44 & 45.55 & 12.92 & 17.41 & 34.43 & 67.17 & 42.59\\
\texttt{SuSimCSE-BERT}        & 12.37 & 18.58 & 39.76 & 68.78 & 44.93 & 13.33 & 16.95 & 34.83 & 67.83 & 43.45\\
\textsc{specter}                & {18.62} & {23.78} & {52.72} & {75.47} & 56.67 & {18.29} & {23.97} & {50.14} & {73.30} & 53.28\\
\textsc{specter-id}             & $\underset{\pm0.92}{\bf 20.09}$ & $\underset{\pm0.45}{\bf 27.36}$ & $\underset{\pm3.04}{\bf 58.74}$ & $\underset{\pm0.41}{\bf 76.49}$ & $\underset{\pm1.31}{\bf 60.40}$ &  $\underset{\pm0.79}{\bf 18.32}$ & $\underset{\pm0.22}{\bf 25.74}$ & $\underset{\pm1.54}{\bf 52.12}$ & $\underset{\pm0.06}{\bf 74.96}$ & $\underset{\pm0.70}{\bf 57.22}$
\end{tabular}   
\label{tab-baseline-results-extended}}
\end{table*}
While Section \ref{sec-baseline-res} presents NDCG$_{\%20}$, we additionally report NDCG$_{\%100}$ in extended results. NDCG$_{\%100}$ indicates a metric comuted based on the entire pool per query. We note based on \citet{wang13ndcg}, that larger pools cause larger values of NDCG, this is observed here. Further model performance at lower values of k, i.e. at the top of the predicted rankings, is still lower indicating significant room for improvements. Finally, note that an apt value of k in computing metrics for evaluation will depend on the choice of target application, we believe trends highlighted in our results hold across values of k as per the consistency of relative performance of models on NDCG$_{\%20}$ and NDCG$_{\%100}$. 

\section{Error Analysis}
\label{appendix-analysis}
Based on a qualitative examination of per-query ranking performance of \texttt{abs\_tfidf, SentBERT-PP} and \textsc{specter-id} we outline a range of factors which lead the baseline models to underperform. We believe the incorporation of modeling to handle these phenomena will lead to improved performance on our dataset. We indicate various error cases through examples of the query facet, false positive top retrievals (FP), or false negative lower ranked retrievals (FN). We mention the query ID for examples in superscripts, use underlines to emphasize important segments, and we only provide the relevant   sentences from the abstract in each example due to space constraints.
 
\textbf{Salient Aspects:} 
One source of error is the inability of models to identify the most salient aspects for similarity, often expressed only in part of a larger set of facet sentences.
\begin{description}[nolistsep, noitemsep]
    \item [\texttt{background Q:}] ``Many classification problems require decisions among a \ul{large number of competing classes}.''$^{1791179}$
    \item \textit{FP:} ``Several real problems involve the classification of data into categories or classes.''
    \item [\texttt{background Q:}] ``With the increasing empirical success of distributional models of compositional semantics, it is timely to consider the types of textual logic that such models are capable of capturing. In this paper, we address \ul{shortcomings in the ability of current models to capture logical operations such as negation.}''$^{1936997}$
\end{description}
Nearly all models miss the notion of negation in the above example.

\textbf{Multiple Aspects: \label{sec-multi-aspect}} 
Within a given facet, papers often expressed multiple finer grained aspects, models however often only retrieved based on a single aspect. In the following baseline models often retrieved based on one or the other aspect:
\begin{description}[nolistsep, noitemsep]
    \item[\texttt{method Q:}] ``We present a Few-Shot Relation Classification Dataset (FewRel), \dots The relation of each sentence is \ul{first recognized by distant supervision methods, and then filtered by crowdworkers}. We adapt the most recent state-of-the-art \ul{few-shot learning methods for relation classification} and conduct a thorough evaluation of these methods.''$^{53080736}$
\end{description}

\textbf{Domain specific similarities} A set of errors also arise from the inability of models to determine similarity between technical concepts. The example represents an inability to rate ``stacking'', ``ensemble strategy'', and ``bagging'' as similar.
\begin{description}[nolistsep, noitemsep]
    \item[\texttt{result Q:}] ``Using a public corpus, we show that \ul{stacking} can improve the efficiency of automatically induced anti-spam filters, \dots''$^{3264891}$
    \item\textit{FN:} ``The experiments on standard WEBSPAM-UK2006 benchmark showed that the \ul{ensemble strategy} can improve the web spam detection performance effectively.''
    \item\textit{FN:} ``We evaluate the classifier performances and find that \ul{BAGGING} performs the best. \dots our method may be an excellent means to classify spam emails''
\end{description}

\textbf{Mechanistic similarities:}
Nearly all methods perform poorly in the case of determining mechanistic similarity in \texttt{method} facets. This often relies on determining similarity across a sequence of actions. Baseline models failed to align steps $^1$ and $^2$ across abstracts below.
\begin{description}[nolistsep, noitemsep]
    \item[\texttt{method Q:}] ``Using an annotated set of"factual"and"feeling"debate forum posts, $^1$we extract patterns that are highly correlated with factual and emotional arguments, and $^2$then apply a bootstrapping methodology to find new patterns in a larger pool of unannotated forum posts.$^{10010426}$''
    \item\textit{FN:} ``$^1$High-precision classifiers label unannotated data to automatically create a large training set, which is then given to an extraction pattern learning algorithm. $^2$The learned patterns are then used to identify more subjective sentences.''
\end{description}

\textbf{Context dependence of facets:}
Faceted similarities as labelled here often also show context dependence on other facets. This is notable in the case of \texttt{result} queries. Given that one major guideline for result similarity in our dataset are if ``the same finding or conclusion'' is found, being able to determine context similarity is important. 
\begin{description}[nolistsep, noitemsep]
    \item [\texttt{result Q:}] ``\dots Subsequently, lexical cue proportions, predicted certainty, \ul{as well as their time course characteristics are used to compute veracity for each rumor tweet} \dots. Evaluated on the data portion for which hand-labeled examples were available, it achieves .74 F1-score on identifying rumor resolving tweets and .76 F1-score on predicting if a rumor is resolved as true or false.$^{5052952}$
    \item \textit{FN:} ``In this study, we propose a novel approach to capture the temporal characteristics of these features based on the \ul{time series of rumor's lifecycle, for which time series modeling technique is applied to incorporate various social context information}. Our experiments using the events in two microblog datasets confirm that the method outperforms state-of-the-art rumor detection approaches by large margins.''
\end{description}
In these examples, determining that the higher level result of time series information being important for identifying rumour tweets relies on modeling method similarity. We believe approaches which improve upon \texttt{method} similarity, will likely benefit overall performance on other facets as well.

\textbf{Qualitative result statements:}
Finally, we also note that \texttt{result} queries which summarize qualitative findings often perform poorer, often requiring broader context and often lacking in term overlaps which may otherwise easily indicate relevance.
\begin{description}[nolistsep, noitemsep]
    \item [\texttt{result Q:}] ``Experiments with several Reddit forums show that style is a better indicator of community identity than topic, even for communities organized around specific topics. Further, there is a positive correlation between the community reception to a contribution and the style similarity to that community, but not so for topic similarity.''$^{11629674}$
\end{description}

\section{Potential training data sources}
\label{appendix-potential-training}
Given these challenges, we also highlight specific other sources of data that future work may exploit to train models to overcome these problems:
\begin{description}[noitemsep, nolistsep]
    \item Domain specific paraphrase datasets: Given the reasonably strong performance of the \texttt{SentBERT-PP} model, fine-tuned on paraphrase datasets, we believe other domain specific paraphrase datasets have the potential to be useful for the proposed task. An example is \textsc{parade} \citep{he2020parade} which presents a dataset of computer science paraphrase pairs.
    \item Selecting informative citation examples: Appendix \ref{appendix-cite-eval} presents an analysis of citation data and indicates how only a part of this data contains fine-grained facet similarities. An potential approach to selecting more informative citation examples might involve model dependent training data subset selection approaches such as that proposed in \citet{antonello2021selecting}.
    \item Co-citations data: Given that the proposed task relies on capturing fine-grained similarities, co-citations examples in the full-text of papers -- papers cited in a narrow context (such as a sentence or paragraph), also promise to contain finer grained similarities likely to help train better models \cite{kobayashi2018citation}. Use of these examples is specially promoted by existence of parsed full-text data in in the S2ORC corpus. 
\end{description}
\end{document}